\documentclass[twocolumn]{aastex631}
\usepackage{graphicx}
\usepackage{float}
\usepackage{breqn}
\usepackage{amsmath}
\usepackage{txfonts}

\accepted{March 7, 2025}

\begin{document}

\title{Photobombing for the Large Interferometer For Exoplanets (LIFE). \\A new criterion for target confusion and application to a MIR rotating nulling interferometer. }

\author{Drinor Cacaj}
\affiliation{ETH Zurich, Institute for Particle Physics \& Astrophysics, Wolfgang-Pauli-Str. 27, 8093 Zurich, Switzerland}
\affiliation{NASA Goddard Space Flight Center, Greenbelt, Maryland 20771, USA}

\author{Daniel Angerhausen}
\affiliation{ETH Zurich, Institute for Particle Physics \& Astrophysics, Wolfgang-Pauli-Str. 27, 8093 Zurich, Switzerland}

\author{Prabal Saxena}
\affiliation{NASA Goddard Space Flight Center, Greenbelt, Maryland 20771, USA}

\author{Romain Laugier}
\affiliation{Institute of Astronomy, KU Leuven, Celestijnenlaan 200D, 3001, Leuven, Belgium }

\author{Jens Kammerer}
\affiliation{European Southern Observatory, Karl-Schwarzschild-Straße 2, 85748 Garching, Germany}

\author{Eleonora Alei}
\affiliation{NASA Goddard Space Flight Center, Greenbelt, Maryland 20771, USA}

\author{Sascha P. Quanz}
\affiliation{ETH Zurich, Institute for Particle Physics \& Astrophysics, Wolfgang-Pauli-Str. 27, 8093 Zurich, Switzerland}
\affiliation{ETH Zurich, Department of Earth Sciences, Sonneggstrasse 5, 8092 Zurich, Switzerland}



\begin{abstract}

One of the primary objectives in modern astronomy is to discover and study planets with characteristics similar to Earth. This pursuit involves analyzing the spectra of exoplanets and searching for biosignatures. Contamination of spectra by nearby objects (e.g., other planets and moons in the same system) is a significant concern and must be addressed for future exo-Earth searching missions. The aim is to estimate, for habitable planets, the probability of spectral contamination by other planets within the same star system. This investigation focuses on the Large Interferometer for Exoplanets (LIFE). Since the Rayleigh criterion is inapplicable to interferometers such as those proposed for LIFE, we present new criteria based on the principle of parsimony, which take into account two types of issues: contamination or blending of point sources, and cancellation of point sources due to destructive interference. We define a new spatial resolution metric associated with contamination or cancellation that generalizes to a broader family of observing instruments. In the current baseline design, LIFE is an X-array architecture nulling interferometer. Our investigation reveals that its transmission map introduces the potential for two point sources to appear as one, even if they do not appear in close proximity. We find that LIFE has a spatial resolution comparable to that of a traditional telescope with a diameter of $D = 600\,\text{m}$, observing at $\lambda = 4 \,\mu\text{m}$. Our survey of a star system population shows that, out of 73.4 expected habitable planets detected, 71.3 are not contaminated on average.

\end{abstract}



\section{Introduction} \label{sec:intro}

\begin{figure}[ht!]
\centering
\includegraphics[width=0.47\textwidth, trim=2cm 4.2cm 2cm 1.8cm,clip]{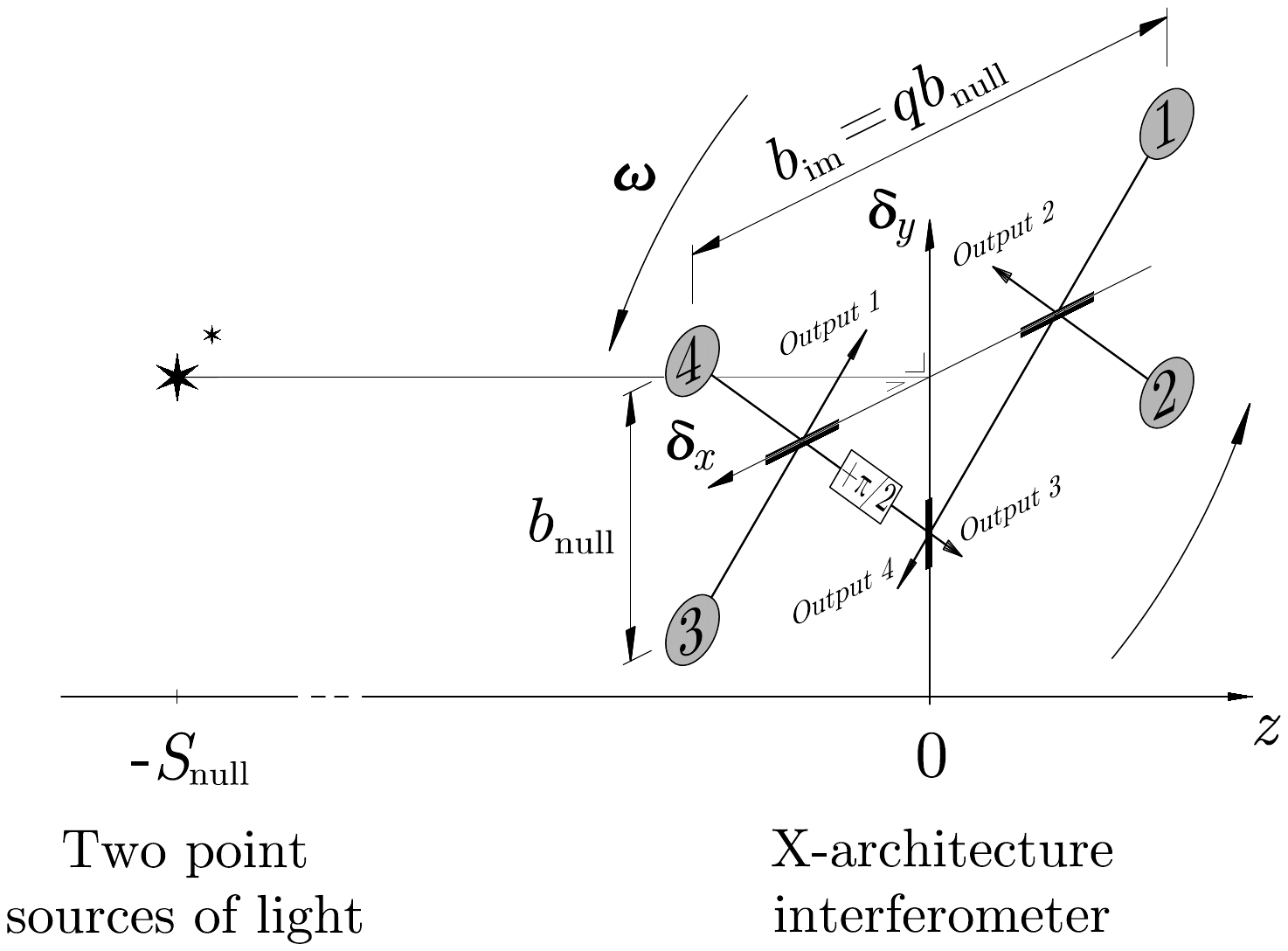}
\caption{X-architecture rotating interferometer setup. LIFE current design has a scaling factor of $q=6$ and nulling baseline $b_{\text{null}}$ that can be adjusted from 10 meters to 100 meters. For details about this setup see \cite{Quanz_2022}. }
\label{LIFE_Setup}
\end{figure}

The Large Interferometer For Exoplanets (LIFE) is a space mission concept aimed at detecting and characterizing exoplanets in the Habitable Zone (HZ) of their host stars (\citeauthor{Quanz_2022}, \citeyear{Quanz_2022}). In its current design, LIFE is a nulling interferometer based on an X-array architecture with four collecting telescopes (Fig.\,\ref{LIFE_Setup}). It has two characteristic dimensions: the nulling baseline $b_{\text{null}}$, which is adjustable from 10 m to 100 m, and the imaging baseline $b_{\text{im}} = 6 b_{\text{null}}$. LIFE observes in the $4\,\mu\text{m}-18.5\,\mu\text{m}$ wavelength range. To make an observation, LIFE must rotate and, unlike traditional telescopes, a monochromatic observation with LIFE will produce a time series instead of an image. A key concept to consider is its ability to resolve two point sources of light that may represent planets.

A first approach to estimate LIFE's spatial resolution is by comparing it to a traditional telescope of diameter equivalent to its imaging baseline. Following \citeauthor{Saxena_2022} (\citeyear{Saxena_2022}), we refer to the terms "photobombing" contamination, or target confusion when two point sources representing planets satisfy a given contamination criterion. In \cite{Saxena_2022}, the Rayleigh criterion was used to estimate spectral contamination. That criterion overestimates spectral contamination. It showed that, for future telescopes of size 6 m and 12 m observing in the NIR, Earth would frequently get contaminated by other planets in the inner solar system, making bio-signatures harder to detect. The aim of this paper is to quantify spectral contamination occurrence for habitable planets by other planets within the same star system. Since, the Rayleigh criterion is not applicable for nulling interferometers we need to find a new contamination criterion that generalizes to traditional telescopes and interferometers like LIFE. This would allow us to compare the resolving capabilities of LIFE with traditional telescopes.

This paper contains key concepts related to the Point Spread Function (PSF) of a traditional telescope and its spatial resolution as defined by the Rayleigh criterion. Applying the principle of parsimony, we attempt to redefine and generalize the concept of spatial resolution by considering two issues that arise when using an observational instrument: the blending of two point sources into one, and the potential for two point sources to cancel each other's signals while remaining detectable independently. The latter phenomenon does not affect traditional telescopes, but it does impact nulling interferometers. We then introduce new metrics denoted by $\delta_0$ and $\delta_1$ that quantifies how much an observing instrument is susceptible to contamination and cancellation respectively. First, we provide the $\delta_1$ spatial resolution for a traditional telescope. In a second step, we examine how LIFE produces time series as images and give its associated $\delta_0$ and $\delta_1$ spatial resolution. The detection of habitable planets is quantified via the detection probability, which is the probability of a given planet being detected. The contamination of habitable planets is quantified via the contamination probability, which is the probability of a given detected planet being contaminated by another planet within the same stellar system. These quantities are calculated assuming that planets do not move during an observation. We briefly present numerical methods in Appendix A to estimate them for two point sources located in the FOV, each with its own spectrum. 

Individual systems like the inner Solar System and the TRAPPIST-1 system are examined. These systems provide two extreme case scenarios for LIFE. The inner solar system can be directly compared with results from \cite{Saxena_2022}. The detection and contamination probabilities will be provided for different values of the system's distance and inclination angle of the orbit plane, followed by a discussion of the results. After this stage, a population analysis study is conducted using the LIFE\textsc{sim} tool (\citeauthor{Dannert_2022} \citeyear{Dannert_2022}) to estimate how much contamination we can expect during the search phase.

\section{Theory}

First, in Sect. \ref{TM1}, we briefly present the current definition of contamination, specifically the angular distance at which two point sources blend or become indistinguishable. In Sect. \ref{TM2}, we introduce a new criterion that is generalized for a broader range of observing instruments, accounting for both contamination and cancellation.

\subsection{Spatial resolution and Rayleigh criterion} \label{TM1}

\begin{figure}[ht]
\centering
\includegraphics[width=0.5\textwidth, trim=2cm 4.2cm 2cm 3.8cm,clip]{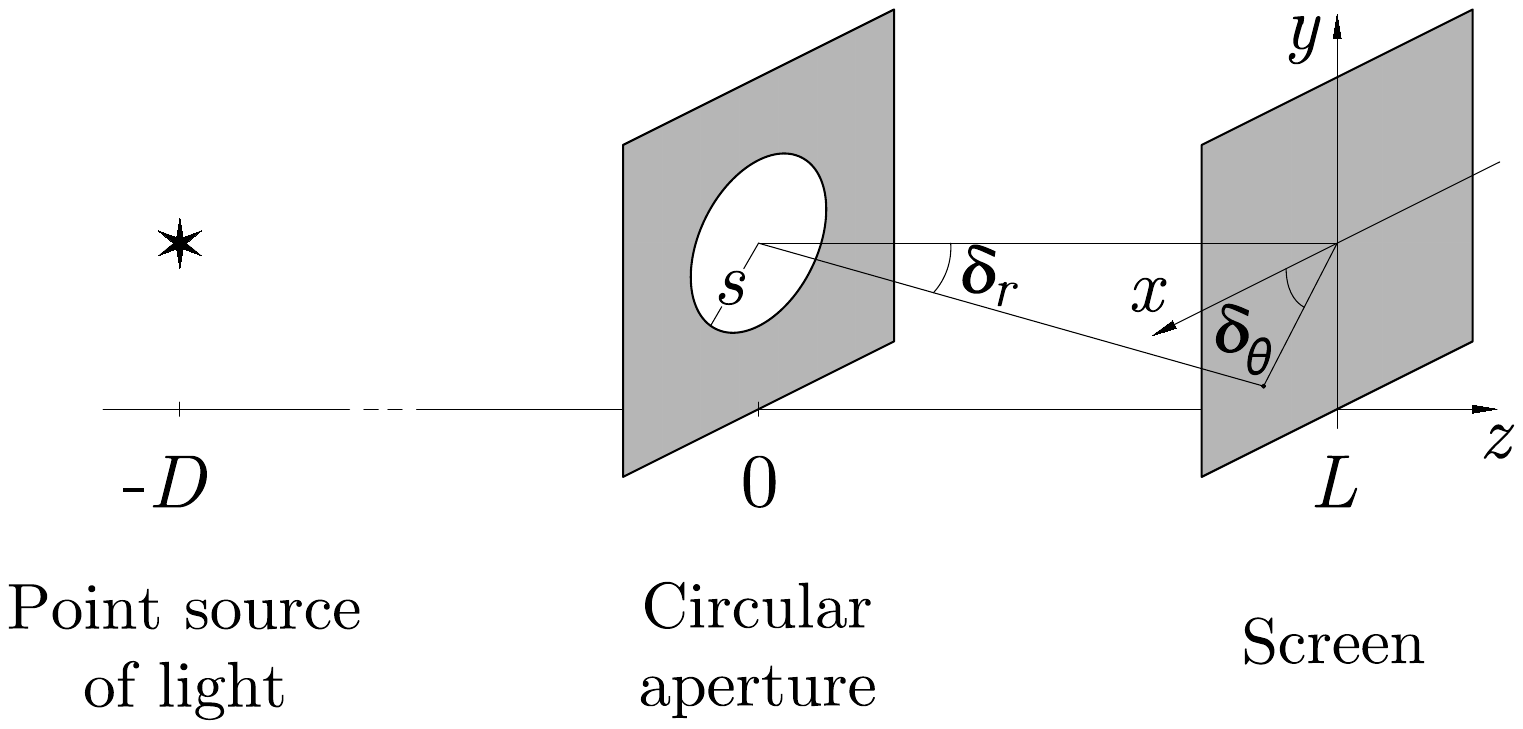}
  \caption{Conventional telescope setup with circular aperture. Far field approximation $D>>L$. Diameter $d = 2s$, where $s$ is the radius of the circular aperture.
          }
     \label{PSF_Setup}
\end{figure}

The PSF describes how a point source of light will appear behind a particular aperture (Fig.\,\ref{PSF_Setup}). To simplify, we will consider a circular aperture and assume that the aperture is in the far field. For a monochromatic point source $p = ((\delta_{x,p}=0, \delta_{y,p}=0), F_p=1)$ where $\delta_{x,p}, \delta_{y,p}$ is the position of the point source in the FOV and $F_p$ is the incoming luminosity flux at wavelength $\lambda$, we get the well-known Airy disk function (\citeauthor{Rayleigh01101879}, \citeyear{Rayleigh01101879}).
\begin{align} \label{Airydiskfunc}
U_p(\delta_{x},\delta_y,\lambda) = \left( \frac{2 J_1( \pi  \delta_rd/\lambda )}{  \pi \delta_rd/\lambda } \right)^2 \qquad \delta_r = \sqrt{\delta_x^2 + \delta_y^2} 
\end{align}
where $\delta_x,\delta_y$ are the angular separation in the $x$ and $y$ axes, respectively, $\lambda$ is the wavelength, $d$ is the diameter of the telescope, and $J_1$ is the Bessel function of the first kind of order 1. The first zero of $J_1(q)$ is at $q = q_0 \approx 3.8317$. The regular notion of spatial resolution is given by the Rayleigh criterion which is defined by the angular separation to the first zero of the Airy disk,
\begin{equation}
\pi \delta_{\text{Rayleigh}} d /\lambda = q_0 \quad \Rightarrow \quad \delta_{\text{Rayleigh}} \approx 1.22\cdot\frac{\lambda}{d} \label{Rayleigh}
\end{equation}
For two monochromatic point sources of light, we say that photobombing/contamination occurs when the apparent angle between them is less than $\delta_{\text{Rayleigh}}$. 

In reality, we can resolve point sources even if their apparent angular separation is smaller than $\delta_{\text{Rayleigh}}$, provided that we have a high enough signal-to-noise ratio (S/N). In theory, in the limit of high S/N, we can always resolve/deconvolute two point sources that do not have the same apparent position. Rayleigh's criterion focuses more on the potential for contamination rather than contamination in an absolute sense. A helpful interpretation is that Rayleigh's criterion intuitively tells us that if two point sources are separated by more than approximately $\delta_{\text{Rayleigh}}$, regardless of the signal-to-noise ratio (S/N), they cannot be mistaken for a single point source. This implies that if the amplitude of the background noise could be adjusted, the two point sources would fade into the noise before merging into one, as long as their angular separation exceeds $\delta_{\text{Rayleigh}}$. It is arbitrarily defined using the first zero of the Airy disk and is therefore ill defined, making it difficult to generalize to other observing instruments.
\subsection{Generalized spectral contamination criterion} \label{TM2}

In this section, we construct a simple criterion to decide if spectral contamination occurs when observing planets within a star system. Assume the ground truth consists of two point sources representing two planets within the a star system. The main idea used in this paper is to compare the best models against the ground truth, factoring in a parsimony cost that increases with the model's complexity. The parsimony cost allows us to select the best model with respect to the number of point sources. If the best model is a model with a single point source then to validate spectral contamination, we add the following key condition: if the best single point source model does not resemble one of the ground truth's point sources, then contamination occurs. This ensures that the best model does not represent something that exists already. A similarity measure between outputs is necessary for comparisons. Outputs can be either image-like or time-series-like, depending on the type of telescope. We will use the Kullback-Leibler (KL) divergence (\citeauthor{kullback1997information}, \citeyear{kullback1997information}) on Gaussian multivariate distributions, which represent the outputs of a given telescope and observed system factoring in the sources and background noise.

\subsubsection{Photon density distributions}

First, we define a point source of light $p$ to be given by its apparent position and incoming photon flux over the wavelengths of observation $\hat{\Lambda}=\{\lambda_1, \dots, \lambda_{M}\}$ defined by the wavelength range $\Lambda$ and spectral resolution $R$
\begin{align}
    p = (\boldsymbol{\theta}_p=(\delta_x,\delta_y) , \boldsymbol{F}_p = ( F_{p,\lambda_1 }, \dots, F_{p,\lambda_{M} }))
\end{align}
In this paper, a given set of point sources is also referred to as a configuration of point sources. They will represent planets located in the FOV.

We define the Unit Instrument Response Function (UIRF) $U_p$ as the noiseless output of a telescope for a given point source $p = (\boldsymbol{\theta}_p , \boldsymbol{F}_p = \boldsymbol{1})$ with unit luminosity flux accross $\lambda_1,\dots,\lambda_M$. It can be image-like or timeseries-like. Unlike the common PSF, it generally depends on the location $\boldsymbol{\theta}_p = (\delta_x,\delta_y)$ in the FOV. For example, for a circular aperture telescope, the UIRF are given by the Airy disk functions $U_p$ (Eq.\,\ref{Airydiskfunc}), centered at $(\delta_{x,p},\delta_{y,p})$, i.e., $\delta_r = \sqrt{(\delta_x- \delta_{x,p})^2 + (\delta_y- \delta_{y,p})^2}$. $U_p$ depends on geometrical hyper parameters. For example, the diameter $d$ of the telescope would be one such parameters.

Given a point source $p = (\boldsymbol{\theta}_p , \boldsymbol{F}_p) $ and a background noise covariance matrix $\Sigma$, we define the monochromatic photon density distribution $\mathcal{N}_p(\lambda)$ as the following multi-variate gaussian distribution,
\begin{align}
    &\mathcal{N}_p(\lambda) = \mathcal{N}\left(\boldsymbol{\mu}_p(\lambda), \Sigma(\lambda)\right) \quad \mbox{where} \quad \boldsymbol{\mu}_p(\lambda) = F_{p,\lambda} \boldsymbol{U}_p(\lambda)
\end{align}
where $\boldsymbol{U}_p = (U_{p,g_1,\dots,g_k})_{(g_1,\dots,g_k)\in\Gamma}$ is a column vector representing the UIRF of the considered telescope for a given point source $p$. $G = \{g_1,\dots,g_K\}$ denotes the output parameters of that telescope. $G=\{\delta_x, \delta_y\}$ for traditional telescopes and $G=\{\rho\}$ for LIFE, where $\rho$ is the angle of rotation of LIFE. $\Gamma$ denotes the output parameter space grid, which is assumed to be equally spaced. For instance, $\Gamma$ can represent the grid of pixels, or it can represent the discrete set of angles of rotation of LIFE. $\mathcal{N}_P(\lambda)$ represents the distribution of monochromatic images given a single point source $p$ factoring in a background noise via the covariance matrix $\Sigma$. In this study we neglect pixel noise, i.e., we assume that the pixel size is small compared to the variation of the UIRF. Moreover, we assume that planet noise is negligible compared to the background noise. Thus, for configurations with multiple point sources $P = \{p_1,p_2\}$ we simply add the means $\boldsymbol{\mu}_p(\lambda)$ while keeping the same background noise, i.e., $\mu_P = \mu_{p_1}+\mu_{p_2}$ and $\Sigma_P = \Sigma_{p_1} = \Sigma_{p_2} = \Sigma$. 

Unlike for traditional telescopes, the convolution idea with PSFs used in traditional telescopes does not work for LIFE. We notice that the shape of the timeseries $\boldsymbol{U}_p$ generally depends on the position of the point source $p$ in the FOV. We will see that for LIFE, there is generally a different UIRF for different apparent positions of $p$ in the FOV. To calculate the photon density distribution of a point source $p$ for LIFE, i.e., a time series for a fixed wavelength, we ``convolute'' with a non-constant ``PSF'', which in this case is the UIRF of LIFE and calculate the background noise term using LIFEsim (\citeauthor{Dannert_2022}, \citeyear{Dannert_2022}).

\subsubsection{The cost function}

In this paper we use the KL divergence $D_{\text{KL}}$ to measure the similarity between two multi-variate gaussian distributions $\mathcal{N}_P(\lambda)$ and $\mathcal{N}_Q(\lambda)$ representing monochromatic outputs from point source $P$ and $Q$ respectively. It has the following closed form solution (\citeauthor{pardo2018statistical}, \citeyear{pardo2018statistical} )
\begin{align}
      &D_{\text{KL}}\left(\mathcal{N}_P(\lambda) \parallel \mathcal{N}_Q(\lambda) \right) \notag \\ 
      & = \frac{1}{2}\left( \operatorname{tr}\left(\Sigma_Q^{-1}\Sigma_P\right) - K + \left(\boldsymbol{\mu}_Q-\boldsymbol{\mu}_P\right)^t \Sigma_Q^{-1}\left(\boldsymbol{\mu}_Q-\boldsymbol{\mu}_P\right) + \ln\left(\frac{\det\Sigma_Q}{\det\Sigma_P}\right) \right) \notag \\
      & = \frac{1}{2} \left(\boldsymbol{\mu}_Q-\boldsymbol{\mu}_P\right)^t\Sigma^{-1}\left(\boldsymbol{\mu}_Q-\boldsymbol{\mu}_P\right)
\end{align}
where $K = |\boldsymbol{\mu}_P| = |\boldsymbol{\mu}_Q|$ and the second equality holds if the background noise of $P$ and $Q$ are the same, i.e., $\Sigma_P = \Sigma_Q = \Sigma$. It makes this similarity notion symmetrical. Moreover, if the covariance matrix is diagonal constant, i.e., $\Sigma_p(\lambda) = \delta_{g_1,g_1'}\cdots\delta_{g_{K},g_{K}'} \sigma^2$ where $\sigma > 0$, it corresponds to the cost function found in \cite{Dannert_2022}. For the sake of consistency we will use the following notation using $J$ to refer to the KL-divergence between two monochromatic photon density distribution with same covariance matrix $\Sigma$,
\begin{align}
    J(P, Q, \Sigma) := & 2D_{\text{KL}}\left(\;\mathcal{N}\left(\sum_{p\in P}\boldsymbol{\mu}_p, \Sigma\right) \;\parallel\; \mathcal{N}\left(\sum_{q\in Q}\boldsymbol{\mu}_q, \Sigma\right)  \; \right)
\end{align}
where $D_{\text{KL}}$ is multiplied by two to have, $J(P,\emptyset,\sigma \boldsymbol{1}) = \text{S/N}(P)^2$ corresponds to the squared expected signal-to-noise ratio of configuration $P$.

\begin{figure}[ht]
\centering
\vspace{-10pt}
\includegraphics[width=0.115\textwidth]{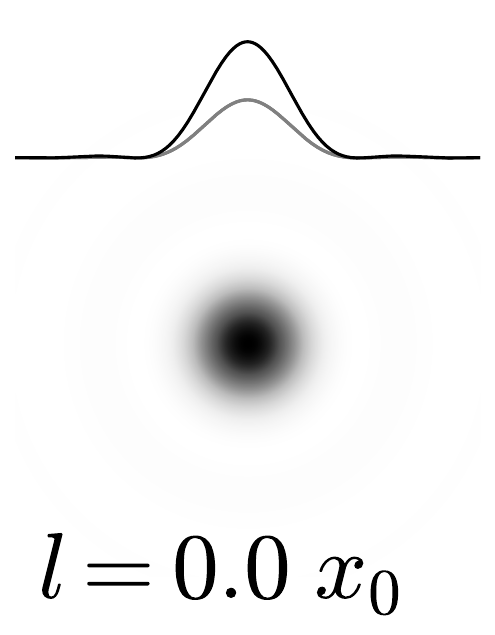}
\includegraphics[width=0.115\textwidth]{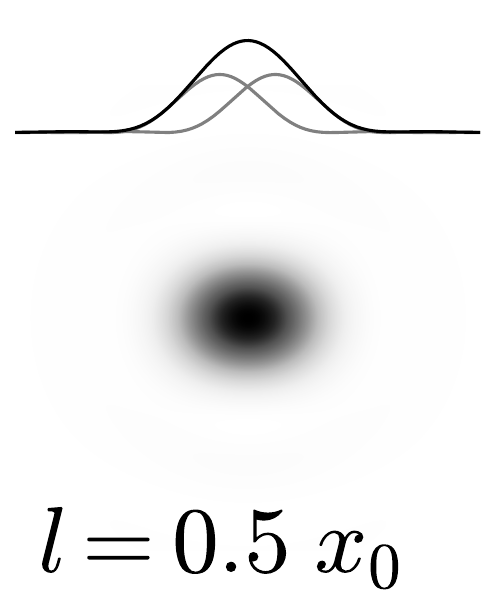}
\includegraphics[width=0.115\textwidth]{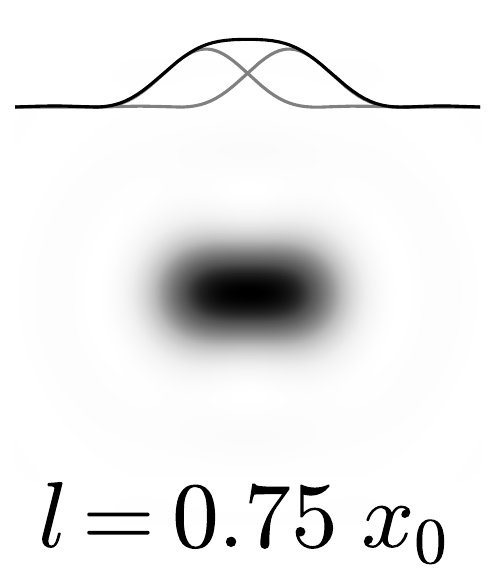}
\includegraphics[width=0.115\textwidth]{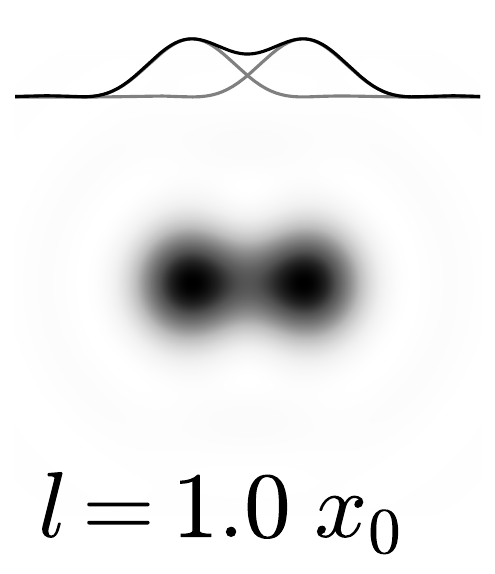}
\caption{Screen outputs from two point sources of light with apparent angular distance between them of $l$ where $x_0 = \delta_{\text{Rayleigh}}$.}
     \label{diffpatterns}
     \vspace{-10pt}
\end{figure}

\begin{figure*}
\centering
\includegraphics[width=0.9\textwidth, trim=0cm 0cm 0cm 0cm,clip]{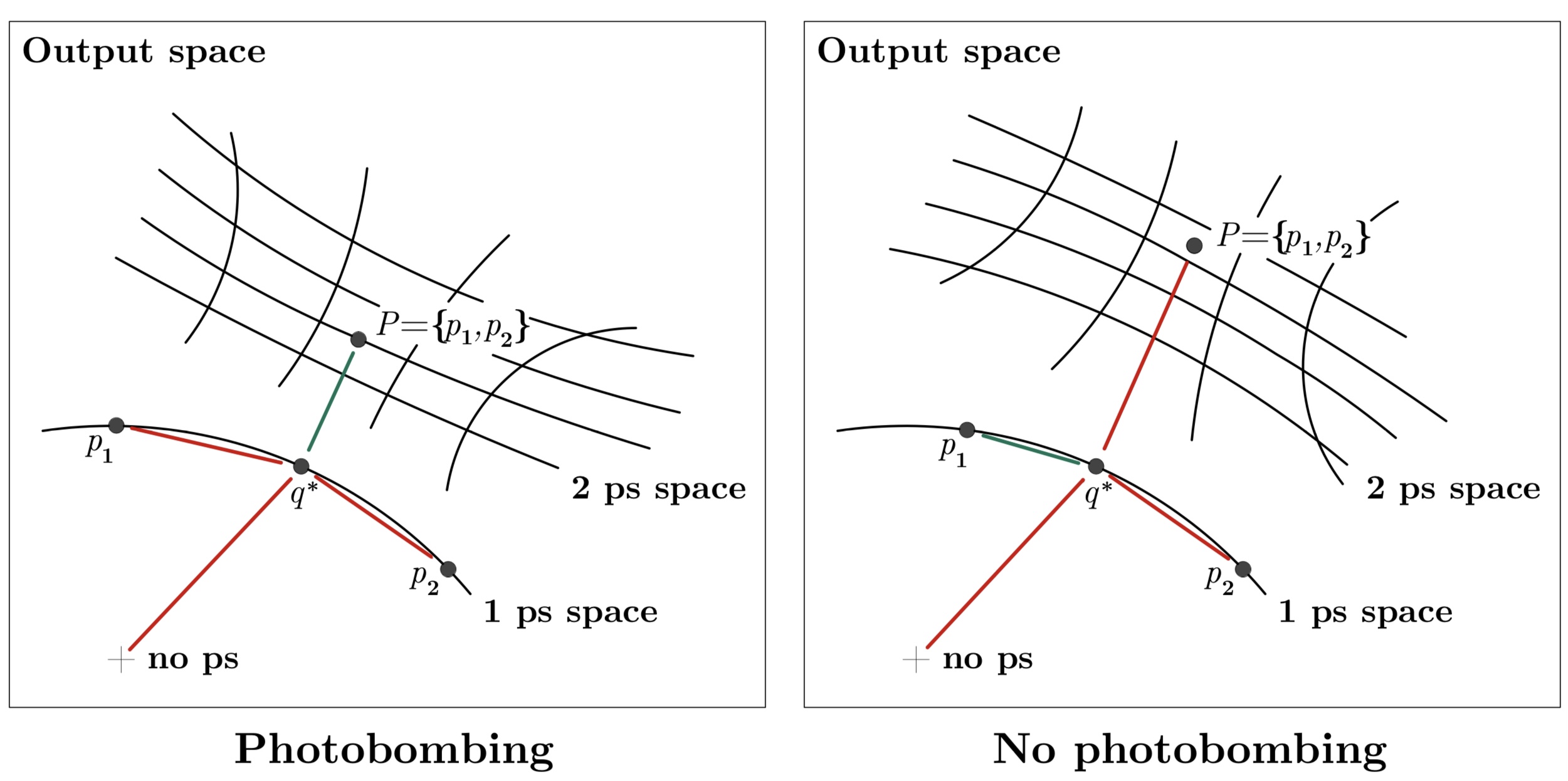}
  \caption{Conceptual examples of photobombing (left) and no photobombing (right) cases. 'Output space' represents all images for traditional telescopes or timeseries for LIFE. 'no ps' represents the output when there is no point source, i.e., a black image or flat timeseries. '1 ps space' represents all noiseless images/timeseries that one fixed point source can produce. Analogously, '2 ps space' represents all noiseless images/timeseries that two fixed point sources can produce. $q^{\star}$ is the 1 point source which minimises the loss with $P$, i.e., P's projection to the generated 1 point source subspace. In green is the smallest distance of the 4 distances according to the similarity notion, which is the loss. If the smallest distance, in green, is between $P$ and $q^{\star}$, then $\mathcal{D}>0$, and by definition, contamination can occurs; otherwise, contamination does not occur. Similar to the Rayleigh criterion, the condition $\mathcal{D}>0$ is more about the possibility and impossibility of contamination rather than contamination in the absolute sense. Notice that 'no ps' $\subset$ '1 ps' $\subset$ '2 ps'. }
     \label{Photobomb_intuition}
\end{figure*}

A few remarks. First, systematic noise affects data points in a similar manner, causing them to deviate together. However, the cost function takes the difference between signal outputs into account, effectively canceling out the systematic noise component. This allows us to focus on the more meaningful variations and patterns in the output. As a result, this cost function provides a measure of similarity that is robust to systematic noise interference. Second, to gain insight into this concept, consider point $P$ in Fig.\,\ref{Photobomb_intuition} as the signal generated by $P$ following an observation, e.g., image with $l=1.0x_0$ in (Fig.\,\ref{diffpatterns}). However, due to background noise, envision a gaussian distribution in this multi-dimensional space of images centered around the mean, which is the noiseless image $\mu_P$ produced by $P$. This multi-variate gaussian distribution illustrates the outcomes of observations, factoring in the noise component. As noise decreases or the S/N increases, the gaussian distribution becomes more localised around $\mu_P$. The quantity $J(P,Q,\Sigma)$ is a proxy for how probable a measure of the output of $P$ would produce the output of $Q$.
\\ \\
A more general definition of target confusion can be expressed using the principle of parsimony. This principle is also known as Occam's razor. Assume we want to explain the output of $P$ with a model $Q_{N}$, which is a configuration of $N$ point sources. Let's add a parsimony cost related to the model's complexity to select the best model. In our case, the model's complexity is described by the number of point sources $N$ of the model $Q_N$. The total cost can be written as follows:
\begin{align*}
    J_{\text{tot}}(P,Q_N,\Sigma) = J(P,Q_N,\Sigma) + \xi(N)
\end{align*}
where $Q_N$ is a configuration of $N$ point sources and $\xi$ is a parsimony cost related to the model's complexity. If $N > M $ then $\xi(N) > \xi(M)$. This model selection requires that $|Q_{\text{tot}}^\star| \leq |P|$, where $Q_{\text{tot}}^\star$ is the model that minimizes the total cost function $J_{\text{tot}}$ with respect to the ground truth $P$. Note that without loss of generality we can set $\xi(0) = 0$. In the next two sections we will study two specific cases where the ground truth is a two point source configuration, which means $|P|=2$.
\subsubsection{Criterion for spectral contamination} \label{specContam}
In this section we study how two point sources can look like one point source that is different from the ones present in the ground truth. Let $|P|=2$ and assume that the model with a single point source $Q^\star = \text{argmin}_{Q:|Q|=1} J(P,Q)$ minimizes the total cost. It implies that the two following equations must be satisfied,
\begin{align}
    &J(P,Q^\star,\Sigma) < \xi(2) - \xi(1) =: \eta^2  \quad &\eta >0 \label{highSNR} \\
    &J(P,\emptyset,\Sigma) > J(P,Q^\star,\Sigma) + \xi(1) > \xi(1) =: \eta_{\text{S/N}}^2 \quad &\eta_{\text{S/N}}>0 
    \label{cond_detection}
\end{align}
The first equation tells us that $Q^\star$ must have a high enough probability of producing an output similar to $P$. We notice that given $\eta$, if $J(P,Q^\star,\Sigma)\neq0$, there is always a scaling of the noise $\beta>0$, ($\Sigma \rightarrow \beta\Sigma$) such that Eq. \ref{highSNR} is not satisfied. It means that there is always a S/N of $P$ such that we can resolve non-overlapping point sources. The second equation tells us that the S/N of $P$ must be higher than a threshold value. It means that there is no possible contamination if there is no detection. Finally, to have contamination, the model $Q^\star$ must be different enough from existing point sources in the ground truth $P$. Indeed, this ensures that the point source described by the simpler model $Q^\star$ does not exist already. For instance, if an observation yields only one detected point source where there are actually two, but the detected point source resembles one of the actual sources, then we have detected something that exists regardless. Hence, we do not consider that as contamination. To have spectral contamination, we impose the following condition:
\begin{equation} \label{alphadef}
    \min_{P_s \in \mathcal{P}(P)-\{P\}} J(P_s,Q^\star,\Sigma) > \alpha\eta^2 
\end{equation}
where $\mathcal{P}(P)=\{ \emptyset,p_1,p_2,P\}$ and $\alpha>0$ a constant. The constant $\alpha$ defines how strong we want the latter condition to be. If $\alpha$ is small then the last condition is weaker, i.e., contamination occurs more frequently.

In summary, for $|P|=2$, according to Eq.\,\ref{highSNR},\ref{cond_detection} and \ref{alphadef}, the best single point source model $Q^\star$ must satisfy three conditions for contamination to occur:
\begin{itemize}
    \item  $P$ must be detectable
\begin{equation}
J(P,\emptyset,\Sigma) > J(P,Q^\star,\Sigma) + \eta_{\text{S/N}}^2
\label{condition_detectable}
\end{equation}
    \item $Q^\star$ must be indistinguishable from $P$
\begin{equation}
    J(P,Q^\star,\Sigma) < \eta^2 
    \label{condition_probable}
\end{equation}
    \item The model $Q^\star$ must be different enough from each existing point sources and void
\begin{equation}
    \min_{P_s \in \mathcal{P}(P)-\{P\}} J(P_s,Q^\star,\Sigma) > \alpha\eta^2 
    \label{condition_different}
\end{equation}
\end{itemize}
Let's define the target confusion map $\mathcal{D}$ of a given configuration with two point sources and background noise covariance matrix $\Sigma$ as,
\begin{equation} \label{TCM}
    \mathcal{\mathcal{D}}(P,\Sigma) = \min_{P_{s} \in \mathcal{P}(P)-\{P\}} J(P_{s}, Q^{\star},\Sigma) - J(P,Q^\star,\Sigma) 
\end{equation} 
where again $Q^\star$ is a single point source that minimizes the cost $J(P,Q)$. The target confusion map is a difference of cost functions and thus is related to a ratio of probabilities. Let $\mathbb{P}_1$ be the likelihood that $P$ can produce the output generated by $Q$ and $\mathbb{P}_2$ be the likelihood that $p_1$ can produce the output generated by $Q$. Then, the ratio of probabilities is given by,
\begin{equation}
    \frac{\mathbb{P}_1}{\mathbb{P}_2} \propto \frac{e^{-J_1}}{e^{-J_2}} = \exp\left(J_2 - J_1\right) = \exp\left(\mathcal{D}\right) \label{fraction_probabities}
\end{equation}
where $J_1 = J(P,Q^{\star},\Sigma)$,  $J_2 = J(p_1,Q^{\star},\Sigma)$. In the third equality we used, without loss of generality, that $J(P_1,Q^{\star},\Sigma) = \min_{P_{s} \in \mathcal{P}(P)-\{P\}} J (P_{s}, Q^{\star})$. Note that $\mathcal{D}=0$ implies that $\mathbb{P}_1=\mathbb{P}_2$. \\ \\ 
By contrapositive, the contamination criterion is not satisfied if the following condition is satisfied,
\begin{equation*}
    \mathcal{D}(P,\Sigma) < (\alpha -1)J(P,Q^\star,\Sigma) 
\end{equation*}
First, note that the only value of $\alpha$ that puts a condition on the probability ratio proxy $\mathcal{D}$ independent of $P$ is $\alpha = 1$. Second, if $\xi(2) = 2\xi(1)$, which is the case for the parsimony cost related to the Akaike Information Criterion (AIC) (\citeauthor{1100705}, \citeyear{1100705}), then $\eta = \eta_{\text{S/N}}$. It means that the cost $J(P,Q^\star,\Sigma)$ must be higher than the detection threshold $\eta_{\text{S/N}}^2$ for contamination not to occur. We conclude that two models are indistinguishable if they have a cost less than the detection threshold. Thus, setting $\alpha = 1$ in  Eq.\,\ref{alphadef} makes sense because we want $Q^\star$ to be distinguishable from $p_1$, $p_2$ or $\emptyset$. These are the reasons why we choose $\alpha = 1$ as a parameter for the condition (\ref{condition_different}). We show in the Appendix B what happens if we vary $\alpha$ for traditional telescopes. 

In summary, there is no contamination if one of the following conditions is satisfied,
\begin{align}
    \text{$P$ is undetectable} \quad & \text{S/N}(P,\Sigma) < \eta_{\text{S/N}} \label{cond_SNR_suffisant} \\
    \text{$P$ and $Q^\star$ are distinguishable} \quad & J(P,Q^\star,\Sigma) > \eta_{\text{S/N}} ^2 \label{cond_distinguishable}\\
    \text{target confusion is negative} \quad & \mathcal{D}(P,\Sigma) < 0 \label{cond_target_confusion_map}
\end{align}
where $\text{S/N}(P,\Sigma) = \sqrt{J(P,\emptyset,\Sigma)}$ is the definition of the S/N and $\eta_{\text{S/N}} = \sqrt{\xi(1)} $ is the detection threshold. Notice that the sign of the target confusion map is independent from the S/N of $P$ and the detection threshold. It means that its value for a given configuration 
$P$ depends only on the family of UIRFs defined by the telescope we are observing with.

Similar to what is presented in \cite{Dannert_2022}, the detection threshold can be set such that the probability that model selection on noise only outputs fails to reject $Q_1^\star$ is small enough,
\begin{align*}
    &\mathbb{P}\left( \text{S/N}\left( Q_1^\star(\boldsymbol{X}), \Sigma \right) > \eta_{\text{S/N}} \right) = \Phi(5) \\
    &Q_1^\star(\boldsymbol{X}) = \text{argmin}_{Q:|Q|=1} \left(\boldsymbol{\mu}_Q-\boldsymbol{X}\right)^t\Sigma^{-1} \left(\boldsymbol{\mu}_Q-\boldsymbol{X}\right)
\end{align*}
where $\boldsymbol{X} \propto \mathcal{N}(\boldsymbol{0},\Sigma)$ and $\Phi(5) \simeq 1-0.9999994$ is the 5-$\sigma$ confidence level. The detection threshold can be approximated using a numerical method that finds $Q^\star$ on a high number of simulated noise only outputs. In this paper, we focus more on the target confusion map criterion (Eq. \ref{cond_target_confusion_map}).  Recall that if $\mathcal{D}>0$ then it is possible but not guaranteed to have contamination. Fig. \ref{Photobomb_intuition} allows us to better see the difference between photobombing (possibility) and non-photobombing cases via the target confusion map. 

A few remarks. First, the condition $J(P,Q^\star,\Sigma) < \eta^2$, where $\eta>0$ is a threshold number, is useful for high S/N scenarios. Specifically, when confronted with a high S/N image of two point sources, as depicted in Fig. \ref{diffpatterns}, we can effectively distinguish point sources even if they are separated by an angle equal to a fraction of $\delta_{\text{Rayleigh}}$. As said in Sect. 2.1, in theory, in the limit of high S/N, we can solve all two point sources if they do not overlap. Throughout this paper, we refer to contamination/photobombing when $\mathcal{D}>0$, yet in practice, we are addressing the possibility of contamination, which is similar to the Rayleigh criterion. Second, by adding the same point source $b$ to $P$ and $Q^\star$ does not change the cost function. Indeed, $b$ can be considered as systematic noise. Thus, for a configuration $P$ of $N$ point sources, we only need to check spectral contamination for each pair of point sources, which is again similar to the Rayleigh criterion used in \cite{Saxena_2022}. Third, the contamination criterion is based on the hypothesis that point sources do not move during an observation. This can be an issue if the observation time is long compared to the orbital dynamics of the system we observe. It is especially true for low mass stars. To include moving targets using the same framework, we need to expand the output space to include all outputs that two moving point sources produce. In other words, we need to include all images or time series that two moving point sources can produce. We also need to redefine what is the model's complexity to use the principle of parsimony and find new algorithms to find $Q^\star$. A first step can be to include all Keplerian moving point sources. This would increase the number of free parameters per point source by 4.

\subsubsection{Criterion for cancelling point sources}
In this section, we study how two point sources can cancel each other's signals, whereas they would have been detected if observed independently. Note that cancellation does not affect traditional telescopes. Let $|P|=2$ and assume that the model with no point source $\emptyset$ minimizes the total cost. It implies that the two following equations must be satisfied,
\begin{align}
    &J(P,\emptyset,\Sigma) <  J(P,Q^\star,\Sigma) + \xi(1)  \\
    &J(P,\emptyset,\Sigma) < \xi(2)
\end{align}
Both the first and second equation tell us that the S/N of $P$ must be lower than a threshold value. We notice that there is no cancellation if the S/N of $P$ is over $\xi(2)$. For cancellation to occur, one or both point sources in $P$ must be detectable when observed independently. Thus, we impose the following condition,
\begin{equation} \label{p1p2detectable}
    \min_{P_s \in \{p_1,p_2\}} J(P_s,\emptyset,\Sigma) < \eta_{\text{S/N}}^2
\end{equation}
Similarly to the previous section we can construct a target cancellation map $\mathcal{C}$ where we assume again that $2\xi(1) = \xi(2)$. Recall that this is the case for the parsimony cost related to AIC and that it implies $\eta = \eta_{\text{S/N}}$. The target cancellation map is defined as follows,
\begin{equation} \label{TcancelM}
    \mathcal{\mathcal{C}}(P,\Sigma) = \min_{P_{s} \in \{p_1,p_2\}} J(P_{s}, \emptyset,\Sigma) - J(P,\emptyset,\Sigma) + J(P,Q^\star, \Sigma)
\end{equation} 
In summary, there is no cancellation if one of the following conditions is satisfied,
\begin{align}
    \text{half $P$ is detectable} \quad & \text{S/N}(P,\Sigma)/\sqrt{2} > \eta_{\text{S/N}} \label{cond_SNR_suffisant_half} \\
    \text{both $p_1,p_2$ undetectable} \quad &  \max_{P_{s} \in \{p_1,p_2\}} \text{S/N}(P_{s}, \Sigma) < \eta_{\text{S/N}} \label{cond_bothundetect}\\
    \text{target cancellation is negative} \quad & \mathcal{C}(P,\Sigma) < 0 \label{cond_target_confusion_map}
\end{align}
where $\text{S/N}(P,\Sigma) = \sqrt{J(P,\emptyset,\Sigma)}$ is the definition of the S/N and $\eta_{\text{S/N}} = \sqrt{\xi(1)} $ is the detection threshold. Notice that the sign of the target cancellation map is independent from the S/N of $P$ and the detection threshold $\eta_{\text{S/N}}$. It means that its value for a given configuration $P$ depends only on the family of UIRFs defined by the telescope we are observing with.
\subsubsection{On multichromatic distributions}
More generally we can define the multichromatic photon density distribution $\mathcal{N}_p$ of a point source $p$ as, 
\begin{align*}
    &\mathcal{N}_p = \mathcal{N}(\boldsymbol{\mu}_p,\Sigma_p) \\
    &\boldsymbol{\mu}_p = \begin{pmatrix}
        \boldsymbol{\mu}_p(\lambda_1) \\
        \vdots \\
        \boldsymbol{\mu}_p(\lambda_M)
    \end{pmatrix} \\
    &\Sigma_p = \Sigma_{\lambda} \otimes \Sigma_{G} = \begin{pmatrix}
        \Sigma_p(\lambda_1) & \Sigma_{p,12} & \dots & \Sigma_{p,1M} \\
        \Sigma_{p,21} & \Sigma_p(\lambda_2) & \dots  & \Sigma_{p,2M} \\
        \vdots & \ddots & \ddots & \vdots \\
        \Sigma_{p,M1} & \Sigma_{p,M2} & \dots & \Sigma_p(\lambda_M)
    \end{pmatrix} 
\end{align*}
where $\otimes $ is the kronecker product, $\Sigma_\lambda$ the covariance matrix related to different observed wavelenghs and $\Sigma_{G}$ the covariance matrix related to different rotation angle $\rho$ for LIFE, or pixels $\delta_x,\delta_y$ for a traditional telescope. The same reasoning as in the previous section can be applied for multichromatic photon density distributions. If we assume that the covariance matrix $\Sigma_p$ is diagonal constant by block, which means that there is no correlation between the flux for different wavelengths of observation, then we only have to replace the cost $J$ by the sum of costs over each $\lambda_k$ factoring in the corresponding noise term $\Sigma(\lambda_k)$. In this paper we will compute the similarity between multichromatic photon density distributions assuming white gaussian noise between images of different colour, i.e.,diagonal constant by block covariance matrices. Under these restrictions, from equation \ref{condition_detectable}, we have that if two planets are detectable at $\lambda_1$ and $\lambda_2$, then they will be detectable at $\{\lambda_1,\lambda_2\}$. Conversely, from equation \ref{condition_probable}, if $P$ and $Q^\star$ are indistinguishable at $\{\lambda_1, \lambda_2\}$, then $P$ and $Q^\star$ will be indistinguishable at $ \lambda_1 $ and  $\lambda_2$. Lastly, for UIRFs that are given by the scaling and translation of some function, e.g., the Airy disk for a circular aperture, equation \ref{condition_different} holds for $\{\lambda_1, \lambda_2\}$ if it is satisfied for $\lambda_1$ and $\lambda_2$. Note that every step in Sect. 2.2.2 can be easily generalized to multichromatic photon density distributions. 
\subsubsection{$\delta_0$, $\delta_1$ and $\delta_{\text{C}}$ spatial resolutions}
In this section, we introduce a new metric to quantify how prone a telescope is to spectral contamination and cancellation. Let $P = \{p_1,p_2\}$ be the ground truth. The main idea is to fix $p_1$ at a specific location in the FOV and move $p_2$ around. We then calculate the surface area within the FOV where the contamination criterion is satisfied $D(\{p_1,p_2\}) > 0$. This area serves as a measure of much the telescope is prone to contamination.

First, we define the following indicator function of a 2 point sources configuration $P = \{p_1,p_2\}$,
\begin{align} \label{indicator}
    \chi(P,\Sigma) = 1_{ \mathcal{D}(P,\Sigma)>0 }
\end{align} 
where $\chi$ is the indicator function, which is a unit step function, and $\Sigma$ is a covariance matrix. The quantity $\chi$ is 1 if contamination can occur and 0 if contamination cannot occur. By fixing $p_1$ and moving $p_2$ over the FOV we can quantify the region in the FOV where there is no possible spectral contamination. Formally, we define the resolution map of a point source $p_1$ with respect to a luminosity flux $\boldsymbol{F}$.
\begin{align} \label{resmapfunc}
    \mathcal{R}(p_1,\boldsymbol{F},\Sigma) = \sqrt{ \frac{1}{\pi} \iint_{\text{FOV}} \chi\left(\left\{p_1,\left((\delta_x,\delta_y),\boldsymbol{F}\right)\right\},\Sigma\right) d\delta_{x}d\delta_{y} }
\end{align}
Notice that if the area of contamination is a disk, then $\mathcal{R}$ represents the radius of the disk. \\ \\
We define the $\delta_1$ resolution at point ($\delta_{x}$, $\delta_{y}$) for a diagonal constant covariance matrix, i.e.,$\Sigma = \sigma \boldsymbol{1}$,
\begin{align}
    \delta_1(\delta_{x},\delta_{y},\hat{\Lambda}) = \mathcal{R}(p,F,\sigma\boldsymbol{1})
\end{align}
where $p = \{\boldsymbol{\theta}_p=(\delta_{x},\delta_{y}), \boldsymbol{F}_p=F\boldsymbol{1}\}$ is a point source with a constant spectrum for $\lambda \in \hat{\Lambda} = \{\lambda_1,\dots,\lambda_M\}$. Since $H(x) = H(cx)$ for any $c >0$, we have independence from $F$ and $\sigma$. $\delta_1$ is only computed to compare different telescope resolving capabilities in a theoretical scenario. That particular scenario is chosen because $\delta_1$ does not depend on signal amplitude $F$ and the noise $\sigma$ amplitude. Notice that $\delta_1$ only depends on the telescopes architecture and thus allows for apple-to-apple comparison between telescopes. In practice, to decide if spectral contamination is possible we check if one of the three conditions (\ref{cond_SNR_suffisant},\ref{cond_distinguishable},\ref{cond_target_confusion_map}) is satisfied for a given configuration of point source $P=\{p_1,p_2\}$ representing planets and covariance matrix $\Sigma$ representing the background noise. \\ \\
A few remarks. First, if $\Sigma = \sigma\boldsymbol{1}$ then the resolution $\mathcal{R}$ depends on the S/N of $p_1$ and $p_2$ via the incoming photon flux $F_{p_1}$ and $F_{p_2}$, and the noise $\sigma$. Using the fact that multiplying the flux of $p_1$ and $p_2$ by a constant $c > 0$ does not change the resolution, we have that $\mathcal{R}$ only depends on the shape of the UIRF. Second, given $p_1$ and a background noise covariance matrix $\Sigma$, the function $f(p_2) = \mathcal{D}(\{p_1,p_2\},\Sigma)$ is bounded by $\mathcal{D}(\{p_1,p_1\},\Sigma) = J(\{p_1\},\emptyset,\Sigma)$ and $-J(\{p_1\},\emptyset,\Sigma)$. Thus, we can normalise the target confusion map as follows, 
\begin{equation}
    \frac{\mathcal{D}(\{p_1,p_2\},\Sigma)}{2J(p_1,\emptyset,\Sigma)} + \frac{\mathcal{D}(\{p_1,p_2\},\Sigma)}{2J(p_2,\emptyset,\Sigma)} \in [-1,1]
\end{equation}
The closer to 1, the harder it is to resolve the point sources. If the value $\mathcal{D}$ is over 0, then it is possible to have spectral contamination. If the value $\mathcal{D}$ is under 0, then there is no spectral contamination. The closer to $-1$, the easier it is to resolve the point sources. Lastly, $\delta_1$ can be generalised for other parameters than $\delta_x,\delta_y$. For example, if we assume Keplerian parameter space for the point sources (6 parameters) we can compute the associated $\delta_1-$resolution.

Similarly, the cancellation map can be normalised as follows
\begin{equation}
    \frac{\mathcal{C}(\{p_1,p_2\},\Sigma)}{2J(p_1,\emptyset,\Sigma)} + \frac{\mathcal{C}(\{p_1,p_2\},\Sigma)}{2J(p_2,\emptyset,\Sigma)} \in [-\infty,1]
\end{equation}
We can define a spatial resolution with respect to the target cancellation criterion. Thus, we define the cancelling spatial resolution $\gamma$ using the same construction, but replacing the criterion in the indicator function (Eq.\,\ref{indicator}) with the target cancelling criterion $\mathcal{C}(P,\Sigma)>0$. 

Lastly, we define the resolution $\delta_{\text{C}}$ associated to the following criterion $\mathcal{C}>0 \cup \mathcal{D}>0$. Note that the surface area in (Eq.\,\ref{resmapfunc}) where $\mathcal{D}>0$ and $\mathcal{C}>0$ are not necessarily exclusive, meaning that $\delta_1 + \delta_0$ does not equal the resolution obtained with the criterion $\mathcal{C}>0 \cup \mathcal{D}>0$. However, we have that
\begin{equation}
    \max(\delta_0,\delta_1) \leq \delta_{\text{C}} \leq \sqrt{\delta_0^2 + \delta_1^2}
\end{equation}
\subsection{Traditional telescope's $\delta_{\text{C}}$-resolution }
\begin{figure*}[ht]
\centering
\includegraphics[width=0.49\textwidth, trim=0cm 0cm 0cm 0cm,clip]{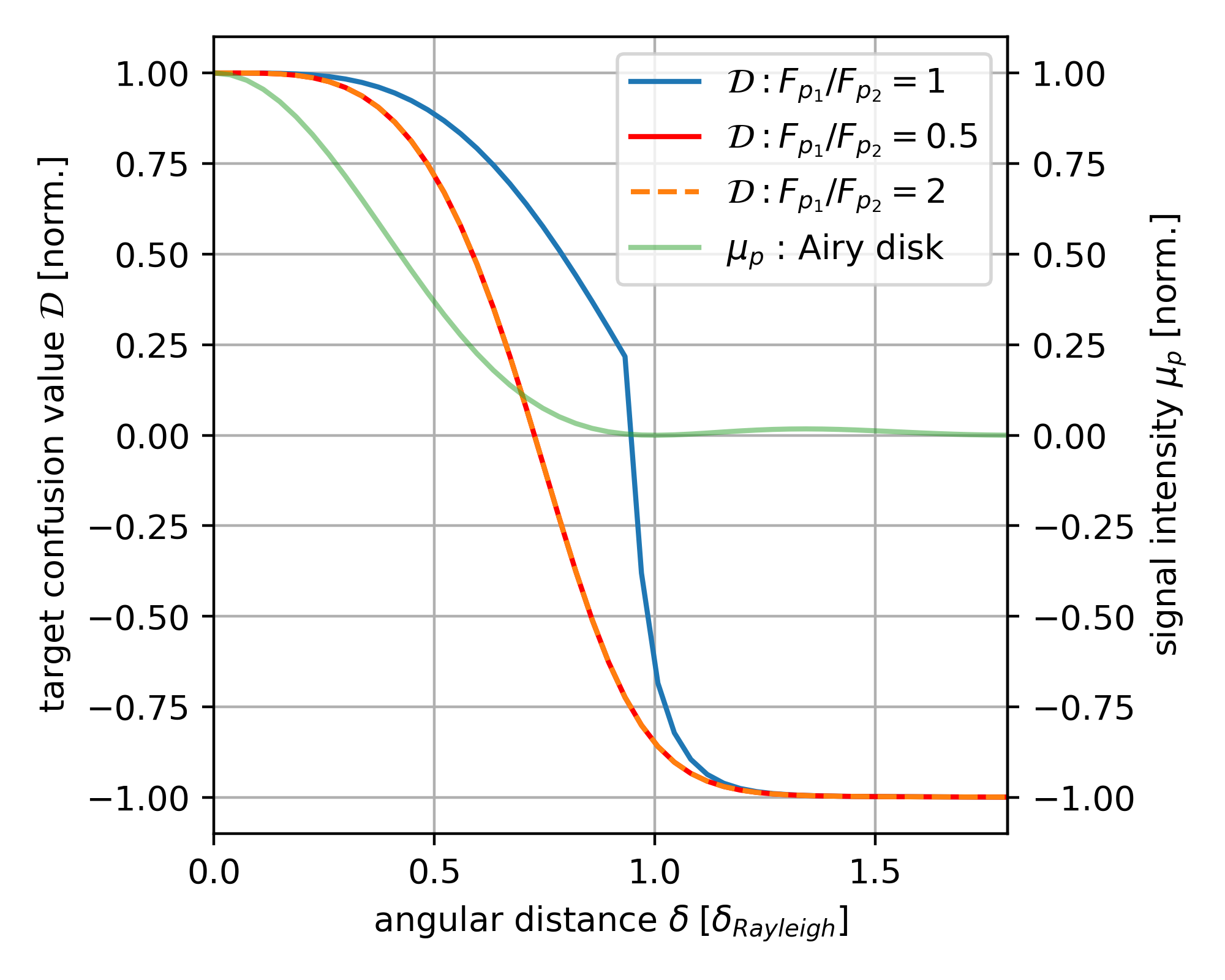}
\includegraphics[width=0.49\textwidth, trim=0cm 0cm 0cm 0cm,clip]{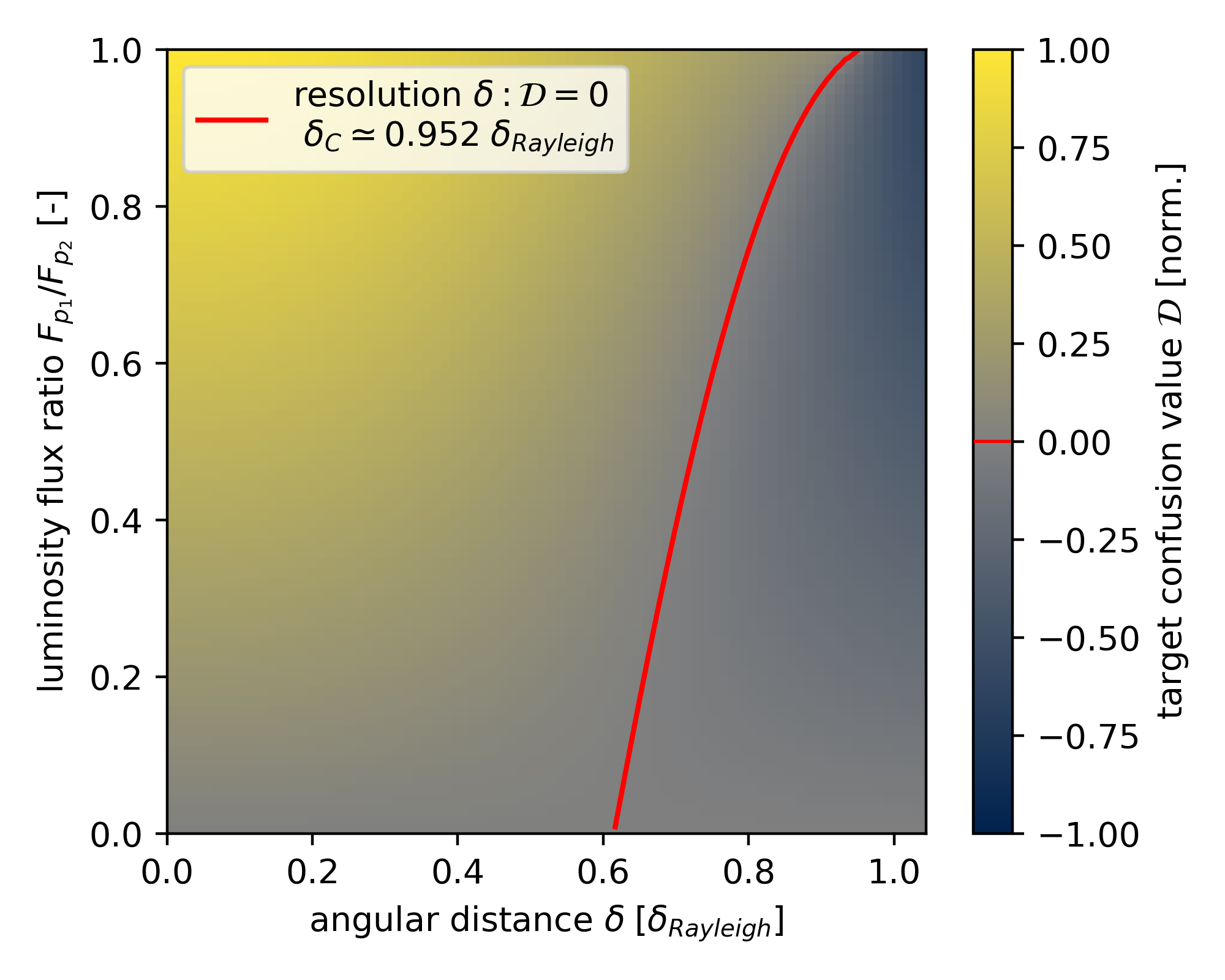}
  \caption{Left : Target confusion map $\mathcal{D}(\{p_1,p_2\},\Sigma)$ where $p_1 = ( \boldsymbol{\theta}_{p_1}= (\delta_{x,1}=0,\delta_{y,1}=0),\boldsymbol{F}_{p_1}=F_{p_1})$ is a monochromatic point source at a fixed position and $p_2 = (\boldsymbol{\theta}_{p_2}=(\delta_{x,2},\delta_{y,2}),\boldsymbol{F}_{p_2}=F_{p_2})$ is a monochromatic point source compared to the Airy disk in light green where $\delta = (\delta_{x,2}^2 + \delta_{y,2}^2)^{1/2}$. $\Sigma$ is diagonal constant covariance matrix representing white background noise. When $p_1$ and $p_2$ are close, the output from $P$ looks like one point source, i.e.,$\mathcal{D}>0$. When far apart the output from $P$ does not look like one point source, i.e.,$\mathcal{D}<0$. $F_{p_1}/F_{p_2}$ is the luminosity flux ratio. Right : Target confusion map $\mathcal{D}(\{p_1,p_2\},\Sigma)$ of a configuration of 2 monochromatic point sources $P = \{p_1,p_2\}$ where $p_1 = (\delta_{x,1}=0,\delta_{y,1}=0,\boldsymbol{F}_{p_1})$ is fixed for different luminosity flux ratios. In red, the resolution map $\mathcal{R}(\{p_1\},F_{p_2})$ for different luminosity flux ratios $F_{p_1}/F_{p_2}$. At $F_{p_1}=F_{p_2}$, $\mathcal{R}=\delta_1$. Spatial resolution $\delta_1 \approx 0.952\,\delta_{\text{Rayleigh}}$ where $\delta_{\text{Rayleigh}}$ refers to the classical notion of resolution, i.e.,first zero of Airy disk. This figure is a generalisation of the plot in the left panel.
          }
     \label{Airyapproach}
\end{figure*}
For a traditional telescope, the only geometrical hyper parameter is the diameter $d$ of the aperture. The output parameters are $g_1 = \delta_x$, $g_2 = \delta_y$, and $\Gamma$ describes the grid of pixels position. The unit output functions are given by the Airy disk (Eq. \ref{Airydiskfunc}) centered at $\boldsymbol{\theta}_p = (\delta_{x,p}$, $\delta_{y,p})$ and the background noise is assumed to be white noise, i.e.,
\begin{align}
    &U_p(\delta_x, \delta_y, \lambda) = \left( \frac{2 J_1( \pi  \delta_rd/\lambda )}{  \pi \delta_rd/\lambda } \right)^2 \\
    & \text{where  } \delta_r = \sqrt{(\delta_x- \delta_{x,p})^2 + (\delta_y- \delta_{y,p})^2} \\
    &\Sigma_p(\lambda) = \delta_{\delta_x,\delta_x'}\delta_{\delta_y,\delta_y'} \sigma_\epsilon^2
\end{align}
where $\sigma_\epsilon$ is the standard deviation of the noise at each pixel position $\delta_x,\delta_y$. Thus, to compute the target confusion value $\mathcal{D}$ for monochromatic photon density distributions we use the following cost function, 
\begin{align*}
    J(P,Q,\Sigma) = \frac{\left(\boldsymbol{\mu}_Q-\boldsymbol{\mu}_P\right)^2}{\sigma^2}
\end{align*}
where $P$ and $Q$ are configurations of point sources and $\sigma = \sqrt{|\Gamma| \sigma_{\epsilon}^2}$ is the integrated standard deviation over the grid $\Gamma$. \\ \\
We show (Fig.\,\ref{Airyapproach}, left) the target confusion map $\mathcal{D}$ for a traditional telescope. Recall that the region of possible contamination is given by $x : \mathcal{D}>0$, where $x$ is the distance between the monochromatic point sources. We notice that the resolution is the same for $F_{p_1}/F_{p_2}$ and $F_{p_2}/F_{p_1}$. This is due to the symmetry of $\mathcal{R}$ coming from $U_p$ (Eq.\,\ref{U_p_LIFE}). The cusp at $\approx 0.9 \delta_{\text{Rayleigh}}$ comes from the fact that the best single point source configuration $Q^{\star}$ is centered at one of the point sources $p_1, p_2$ at greater angular separations, and that it is centered in between if $p_1$ and $p_2$ are close enough. Moreover, we show (Fig.\,\ref{Airyapproach}, right) how the $\delta_1$-resolution changes with the luminosity flux ratio $F_{p_1}/F_{p_2}$. It seems that two point sources with the same luminosity flux are slightly harder to separate than two point sources with different luminosity fluxes. The $\delta_1$-resolution of a traditional telescope computed numerically is:
\begin{align}
    \delta_{\text{C}}(\{\lambda\}) = \delta_1(\{\lambda\}) \approx 0.952 \cdot\delta_{\text{Rayleigh}}(\lambda) \approx 1.16 \cdot \frac{\lambda}{d} 
\end{align}
where we used that $\delta_0 = 0$ for traditional telescopes, meaning that $\delta_{\text{C}} = \delta_1$. Notice that in this case, $\delta_1$ is independent from $(\delta_{x}, \delta_{y})$, because $U_p$ has a constant shape for any position of a point source $p$. We also notice that $\delta_{\text{Rayleigh}}$ is a close approximation of $\delta_1$. A convergence rate plot (Fig.\,\ref{convergence_Airy}) for the $\delta_1$-resolution is presented in the Appendix C. 

A few remarks. First, the Rayleigh criterion was defined using $\delta_{\text{Rayleigh}}$, which was itself arbitrarily defined by the first zero of the Airy disk. The purpose of this angle is not only to compare the spatial resolution capacities of telescopes with each other; i.e., it is not only a standard angle. It is an angle that tells us approximately at what angular separation two point sources can never be confused into only one point source that does not resemble one of the two point sources, regardless of any S/N. Second, $\delta_1$ clearly overestimates contamination because for two point sources where the flux ratio $F_{p_1}/F_{p_2}$ is approaching 0, we still have $\delta_1 > 0$. In these case contamination should be negligible. Lastly, $\gamma = 0$ for traditional telescopes. This is because there is no destructive interference in such instruments.

\subsection{LIFE and nulling baseline optimization}
\begin{figure}[ht]
\centering
\vspace{-10pt}
\includegraphics[width=0.48\textwidth, trim=0.2cm 0.4cm 0cm 0.4cm,clip]{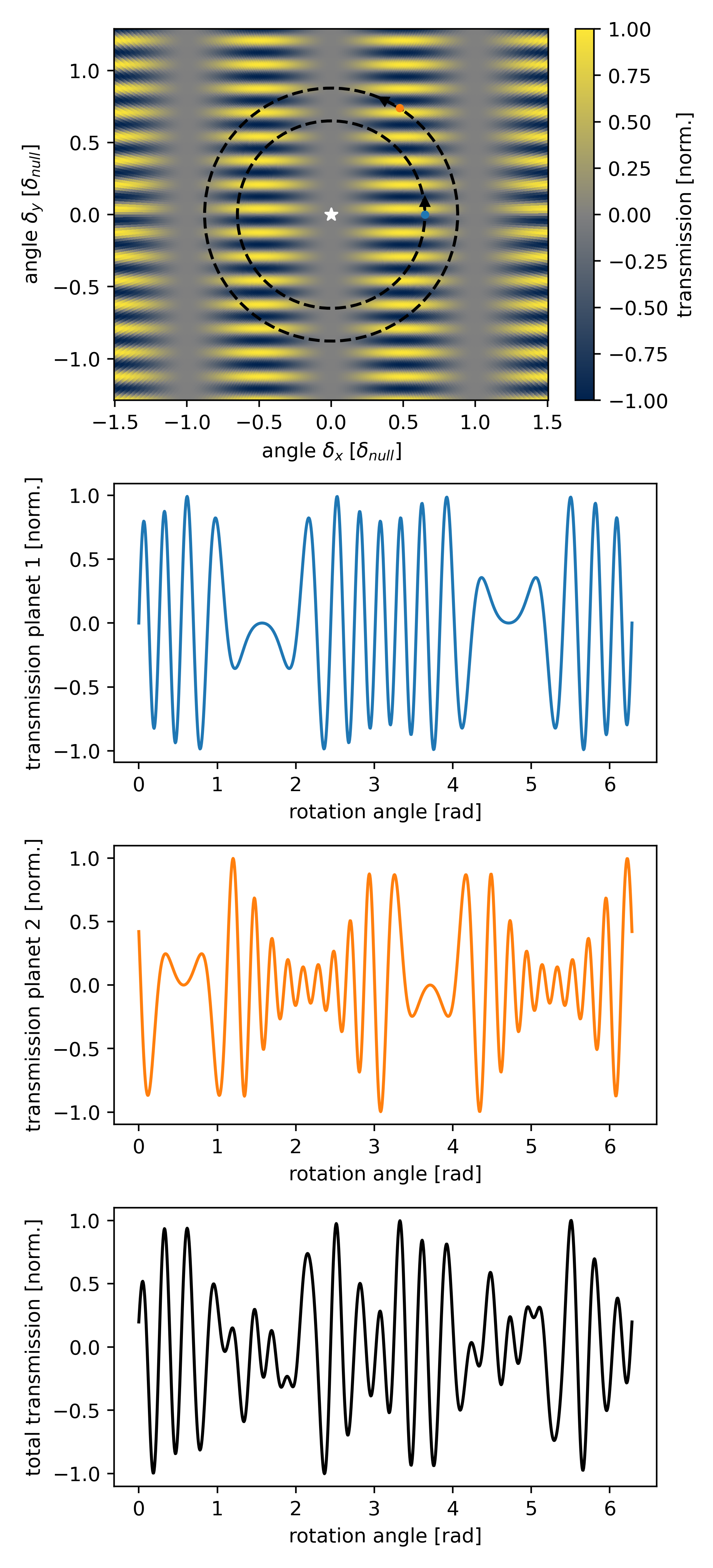}
  \caption{Top : Normalised differential transmission map $T$ at $\lambda = \lambda_0 = 10\,\mu\text{m}$ (Eq.\,\ref{T}) compared to the apparent position of a star at the center and the apparent path of two exoplanets with an apparent angle of $\delta_r = 0.65 \delta_{\text{null}}$ and $\delta_r = 0.88 \delta_{\text{null}}$ from its star after rotating LIFE about LIFE-star axis. Second and third plot are the normalised timeseries of the corresponding planets in the top plot. Bottom : Normalised timeseries of the total signal of the planets in the top plot where planet 1 (blue) has twice the luminosity of planet 2 (orange) at $\lambda = 10\,\mu$m.
          }
     \label{diffmap}
\end{figure}

The Double-Bracewell beam combination scheme currently considered for LIFE's X-array architecture leads to the following transmission map $T$ (\cite{Dannert_2022})
\begin{equation} \label{T}
    T(\delta_r,\delta_{\theta},\lambda) = \sin\left(\frac{\pi b_{\text{null}}}{\lambda} \delta_r\cos(\delta_{\theta})\right)^2\sin\left(\frac{2\pi q b_{\text{null}}}{\lambda} \delta_r\sin(\delta_{\theta})\right)
\end{equation} 
where ($\delta_r, \delta_\theta$) is the apparent position of the planet/point source. We define $\delta_{\text{null}} = \lambda_0/b_{\text{null}}$ where $\lambda_0 = 10\, \mu\text{m}$. \\ \\
For LIFE, the unit output functions are given by timeseries around its transmission map (\ref{T}) and the background noise is assumed to be white noise, i.e. :
\begin{align}
    &U_p(\rho, \lambda) = T(\delta_{r,p}, \delta_{\theta,p} + \rho ,\lambda) \label{U_p_LIFE} \\
    &\Sigma_p(\lambda) = \delta_{\rho\rho'}\sigma_{\epsilon}(\lambda)
\end{align}
where $\rho$ is the rotation angle of LIFE, which is the only output parameter, and $\Gamma = \{ 0,\dots,2\pi\}$. $\sigma_{\epsilon}$ is the standard deviation of the noise at each 'pixel' $\rho \in \Gamma$. To gain more insight, for a fixed $\lambda$, the output signal is in the form of a time series where 'time' is the rotation angle of the array. A point source $p$ with apparent position $\boldsymbol{\theta}_p = (\delta_x,\delta_y) = (\delta_r\cos(\delta_{\theta}),\delta_r\sin(\delta_{\theta}))$ and spectral photon flux $\boldsymbol{F}_{p,\lambda}$ at wavelength $\lambda$ yields the following time series:
\begin{align}
    \mu_p(\rho) = F_{p,\lambda}T(\delta_{r,p}, \delta_{\theta,p} + \rho ,\lambda)
\end{align}
where $\rho$ is the angle of rotation of the array. An example with two time series resulting from two planets located at two different positions is shown in (Fig.\,\ref{diffmap}).\\ \\ 
The modulation efficiency (\cite{Lay:04}) is given by the following,
\begin{align} \label{S}
    \xi(\delta_r) = \sqrt{ \int_{4}^{18.5}\int_{0}^{2\pi}  \mu_p^2(\rho, \lambda) d\rho d\lambda }
\end{align}
and quantifies how sensitive the telescope is for a planet that has an angular separation of $\delta_r$ from the star. It is related to the cost function $\sqrt{J(P=\{p_1\},\emptyset,\Sigma = \boldsymbol{1})}$. 
\\ \\ 
If one does not know the inclination angle of the orbit plane $\phi$ (by definition face-on $ \equiv \;\phi = 0$), the expected signal intensity over the distribution of inclination angles $\phi$ and circular orbit trajectory is also useful to know. In that case, the point source angular separation $\delta_r$ is given by:
\begin{equation}
     \delta_r = \frac{s}{S_{\text{dist}}} \sqrt{\cos^2(\theta) + \sin^2(\theta)\cos^2(\phi)}
\end{equation}
where $s$ is the semi-major axis and $S_{\text{dist}}$ is the observer-star distance. Note that $\phi \propto \arccos(u)$ where $u\in[0,1]$ is uniformly distributed. 
\begin{figure}[ht]
\centering
\vspace{-10pt}
\includegraphics[width=0.5\textwidth, trim=0.2cm 0cm 0.6cm 0cm,clip]{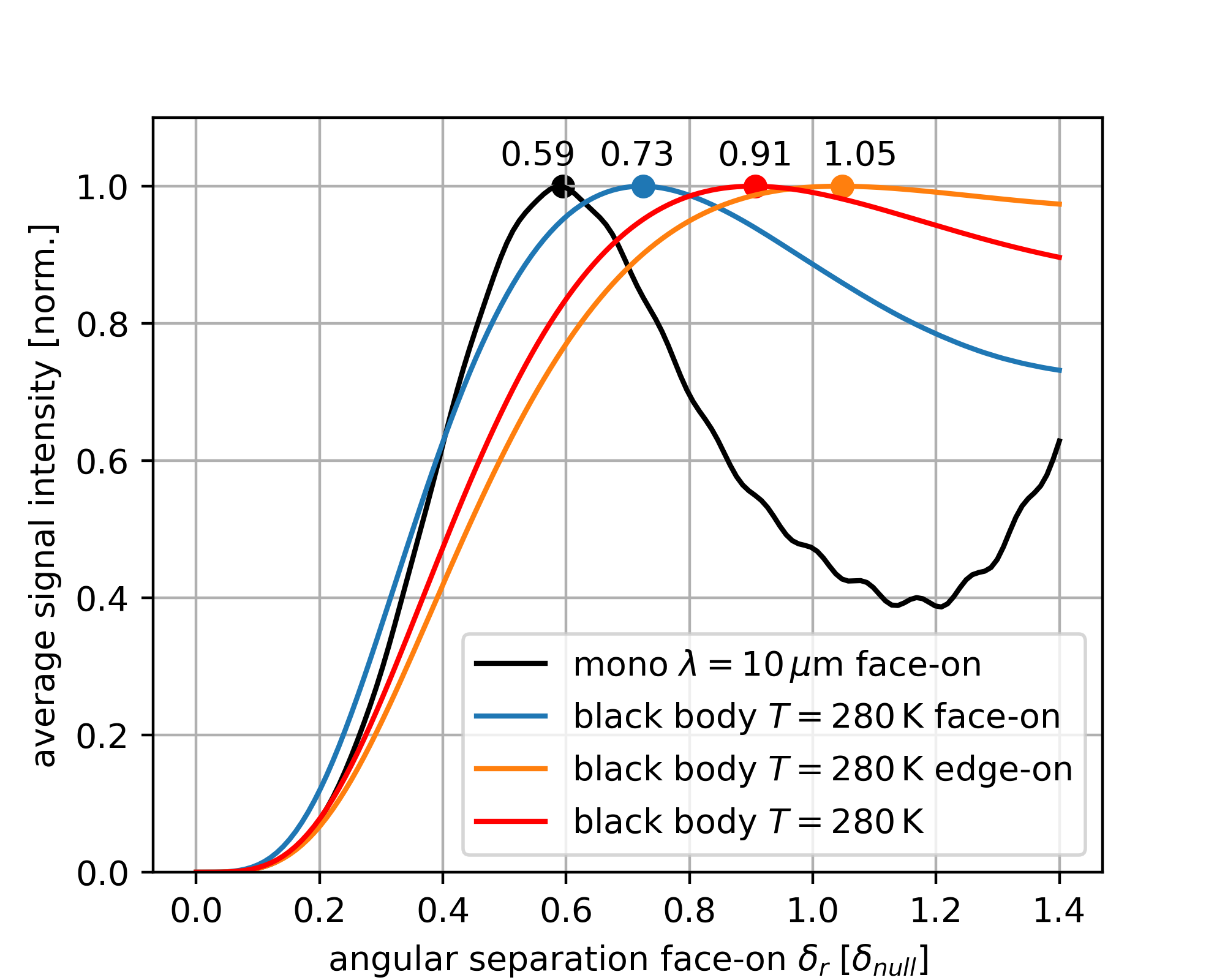}
  \caption{Normalised expected signal intensity for a black body radiating point source at temperature $280\,\text{K}$ with maximum semi-major axis angle of $\delta_r$. Black is the signal intensity at $\lambda = 10\,\mu\text{m}$ and face-on, i.e.,$\phi=0$. Blue is averaged over orbital angle $\theta$ face-on. Orange is averaged over orbital angle $\theta$ edge-on, i.e.,$\phi=\pi/2\,\text{rad}$. Red is averaged over $\theta$ and $\phi$. Without any information about $\phi$ we set the baseline such that the apparent HZ center radius angle $\delta_{HZ}$ matches $0.91\cdot\delta_{\text{null}}$.
          }
     \label{diffmap4}
\end{figure}
By averaging over both $\phi$ and $\theta$ angles and assuming the luminosity flux to be a black body radiance spectrum, we can determine the semi-major axis at which the telescope exhibits the highest average sensitivity for a planet of temperature $T$. We show (Fig.\,\ref{diffmap4}) the normalised signal intensity for a black body radiating point source at Earth's surface temperature $T=280\,\text{K}$. For a planet with temperature $T = 280\,\text{K}$, we find that the planet signal is maximized when $\delta_r = \delta_H :=  0.91\cdot \delta_{\text{null}}$. In our analysis, we will try to set $\delta_{\text{HZ}} = \delta_H$ to maximise the planet's signal. This is not always possible since $b_{\text{im}} < 600\,\text{m}$. For these cases we will set $b_{\text{im}} = 600\,\text{m}$.
\subsection{LIFE's $\delta_0$ and $\delta_1$ spatial resolutions}
The corresponding cost between multichromatic photon density distributions is given by the same cost as found in (\citeauthor{Dannert_2022}, \citeyear{Dannert_2022}), 
\begin{align*}
    J(P,Q,\Sigma) =  \sum_{\lambda_k} \frac{\left(\boldsymbol{\mu}_Q(\lambda_k)-\boldsymbol{\mu}_P(\lambda_k)\right)^2}{\sigma(\lambda_k)^2} \Delta \lambda_k
\end{align*}
where $P$ and $Q$ are configurations of point sources and $\sigma(\lambda_k) = \sqrt{|\Gamma|\sigma_\epsilon(\lambda_k)}$ is the integrated noise over $\Gamma$. In practice, we have access to $\sigma$ via LIFEsim which considers all relevant astrophysical noise terms.

\begin{figure}[ht]
\centering
\includegraphics[width=0.46\textwidth, trim=0.2cm 0.4cm 0.cm 0.3cm,clip]{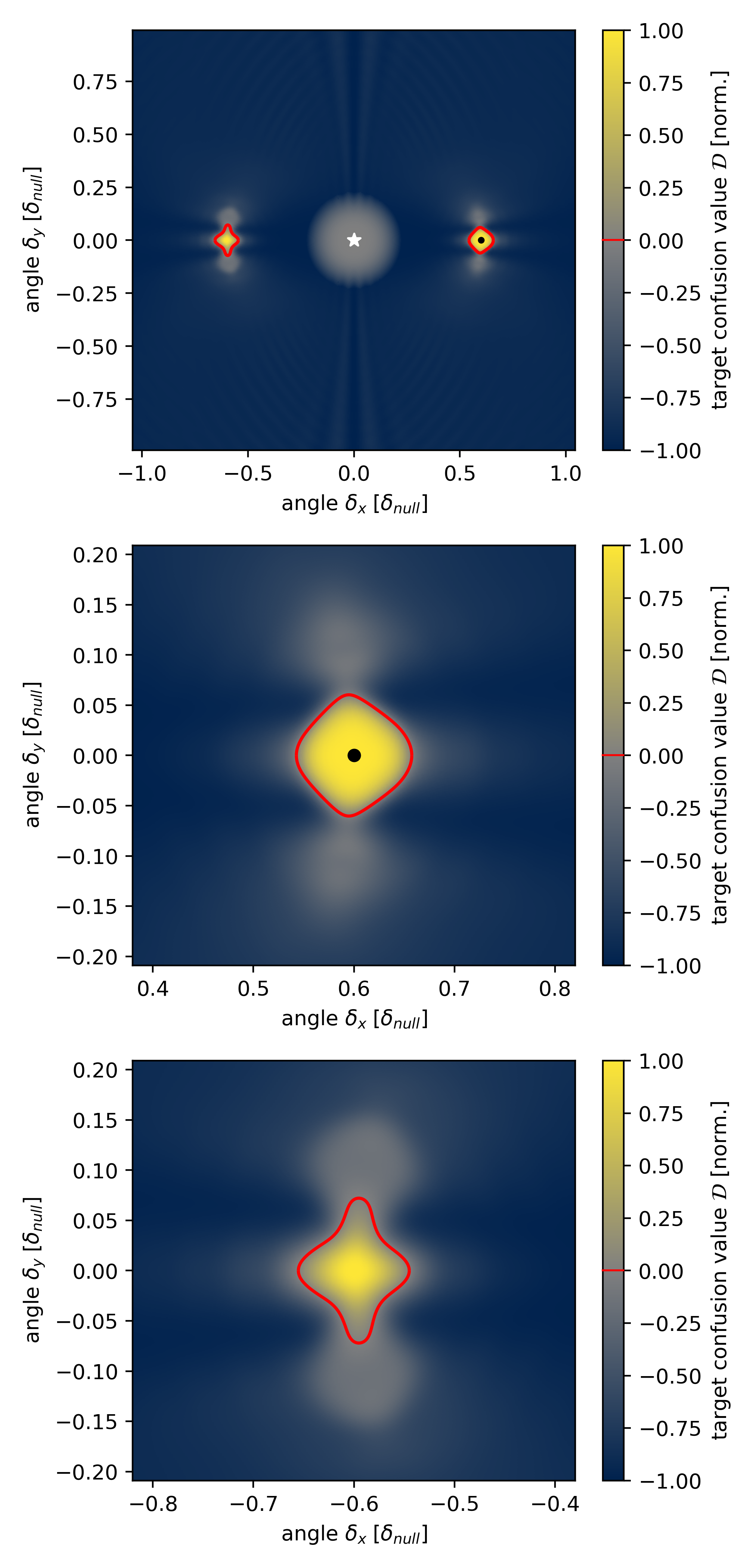}
  \caption{Top : Normalised target confusion map $\mathcal{D}$ (Eq.\,\ref{TCM}) for $p_1 = ((0.6\cdot\delta_{\text{null}}, 0), \boldsymbol{F}_{p_1})$ (black dot) with homogeneous luminosity flux $\boldsymbol{F}_{p_1} = F_{p_1}\boldsymbol{1}$ across the full wavelength range, $\lambda$ from $4\,\mu$m to $18.5\,\mu$m and spectral resolution of 20. At each point ($\delta_x,\delta_y$) is computed the target confusion map evaluated at $P = \{p_1,p_2\}$ where $p_2 = ( (\delta_x,\delta_y), \boldsymbol{F}_{p_2} = 0.5 F_{p_1}\boldsymbol{1} )$ with constant noise independent from $P$, i.e.,$\sigma(\lambda)=\sigma$. When $p_2$ is outside the red region photobombing does not occur. The red region is independent from $F_{p_1}$ and $\sigma$. In this case $b_{\text{im}} = 6\,b_{\text{null}}$ and $\mathcal{R} \approx 0.08\,\delta_{\text{null}}$. (Middle) Zoom in the region of local photobombing, i.e.,2 point sources look like one non-existing point source and they appear in close proximity. Bottom : Zoom in the region of non-local photobombing, i.e.,2 point sources look like one non-existing point source and they do not appear in close proximity. }
  \vspace{-15pt}
     \label{photobombingmap}
\end{figure}

We show (Fig.\,\ref{photobombingmap}) the target confusion map $\mathcal{D}$ (Eq.\,\ref{TCM}) for a configuration of two point sources with one point at a fixed position. We see that there are two regions of interest delimited by the 0 level line in red ($\mathcal{D}=0$). If $p_2$ is inside the region $\mathcal{D}>0$ then photobombing is possible. The resolution $\mathcal{R}$ is defined using the area of this region. If $p_1$ and $p_2$ overlap (black dot), the resulting $Q^\star$ is a point source with luminosity flux $\boldsymbol{F}_{q} = \boldsymbol{F}_{p_1} + \boldsymbol{F}_{p_1}$ located at the same position. If $p_2$ is at the opposite side and $\boldsymbol{F}_{p_2} < \boldsymbol{F}_{p_1} $ then for all $\lambda$, $p_2$ would cancel the signal from $p_1$ in a coherent manner. Thus, the resulting $Q^\star$ is a point source with luminosity flux $\boldsymbol{F}_{q} = \boldsymbol{F}_{p_1} - \boldsymbol{F}_{p_2}$ at $p_1$'s position. Note that these addition and subtraction rules are valid only at 2 positions, namely the same for addition and the opposite for subtraction. More generally, if the point source $p_2$ is in the region $\mathcal{D}>0$, the resulting S/N of $Q^\star$ would be a more complex combination of $p_1$'s and $p_2$'s spectra. We will provide a concrete example in the study of an Earth twin in Sect. 3.1.1. Finding $Q^\star$ is key to studying the occurrence of photobombing. Practically, in this paper, this is done using linear regression and Newton-Raphson method techniques alternatively. We use the analytical expression of the transmission map to make faster and more accurate computations.

\begin{figure*}[ht]
\centering
\includegraphics[width=0.49\textwidth, trim=0.2cm 0cm 0.cm 0cm,clip]{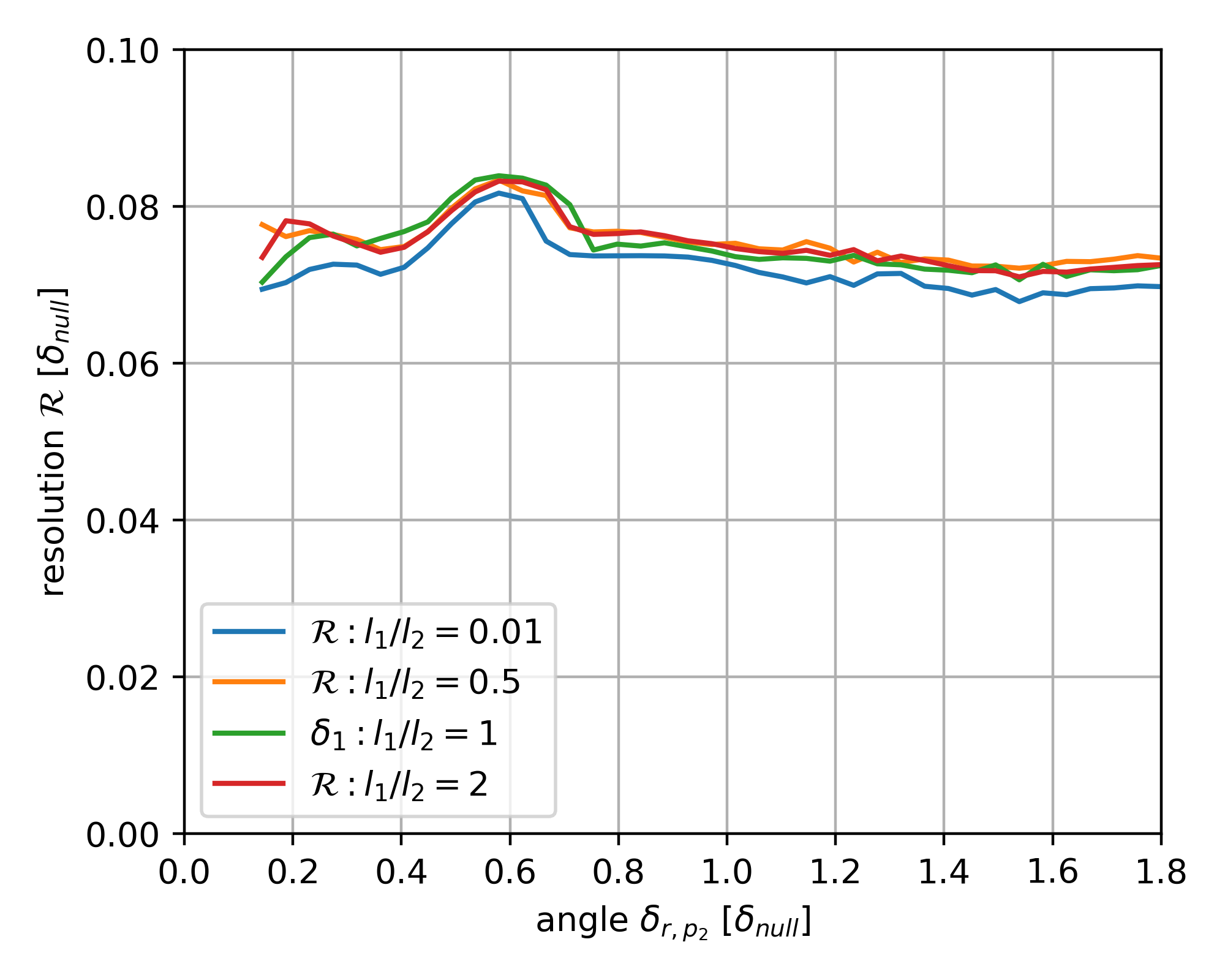}
\includegraphics[width=0.49\textwidth, trim=0.2cm 0cm 0.cm 0cm,clip]{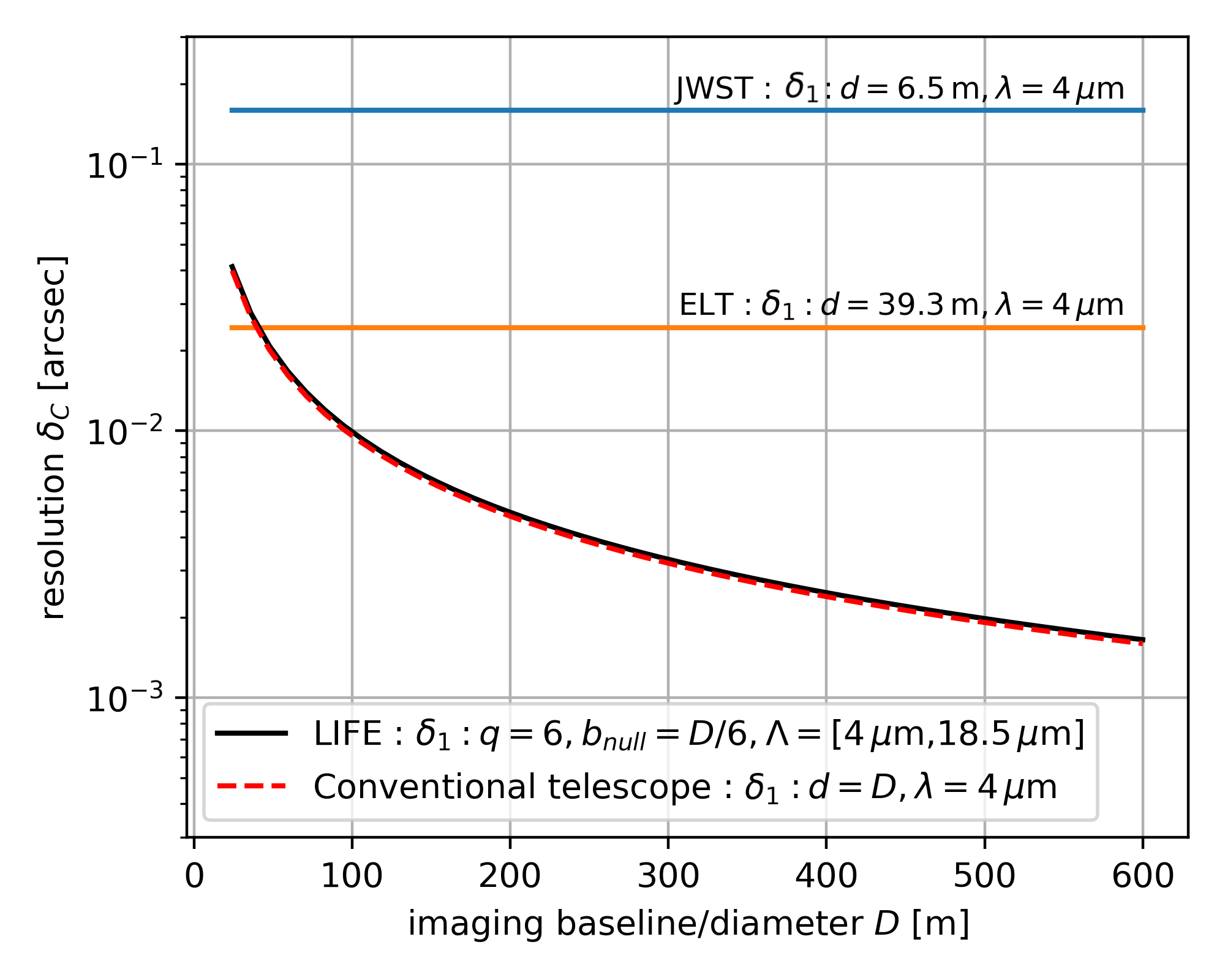}
  \caption{ Left : Resolution $\mathcal{R}(p_1, \boldsymbol{F}_{p_2}=F_2\boldsymbol{1}, \sigma(\lambda)=\sigma)$ (Eq.\,\ref{resmapfunc}) for point sources with homogeneous luminosity fluxes $\boldsymbol{F}_{p_i} = F_{p_i}\boldsymbol{1},\;i=1,2$. At each point $\delta_{r}$ is computed the resolution, i.e.,square root of $1/\pi$ of the area of contamination evaluated at $P = \{p_1,p_2\}$ where $\delta_{r,p_1} = \delta_r$ and $\sigma_{P} = \text{const.}$. $\mathcal{R}$ weakly depends on the luminosity flux ratio $F_r=F_1/F_2$. Right : Comparison between LIFE's $\delta_1$ resolution with JWST and ELT. In this case $b_{\text{im}} = 6\,b_{\text{null}}$. }
     \label{resmap}
\end{figure*}

We show (Fig.\,\ref{resmap}, left) the resolution map $\mathcal{R}(p_1, \boldsymbol{F}_{p_2}=F_{p_2}\boldsymbol{1},\Sigma=\text{diag}_{\lambda_k\in\hat{\Lambda}}(\Sigma(\lambda_k)))$ where $p_1 = (( \delta_{r,p_1} , 0), \boldsymbol{F}_{p_1}=F_{p_1}\boldsymbol{1})$ and $\Sigma(\lambda_k)$ is diagonal constant. We see that the resolution is not highly dependent on the luminosity flux ratio $F_{p_1}/F_{p_2}$ and the angular separation $\delta_{r,p}$ for $\delta_{r,p} > 0.15\, \delta_{\text{null}}$. This map characterises the performance of the transmission map to resolve two point sources and it can be compared to resolution maps from other transmission maps. We notice (fig\,\ref{resmap}, right) that the $\delta_1$-resolution of LIFE is surprisingly similar to the $\delta_1$-resolution of a large traditional telescope of diameter $d=b_{\text{im}}$ observing at $4\,\mu\text{m}$. This exactly corresponds to the first approximation made in the introduction without using this new theoretical framework. 

\begin{figure}[ht]
\centering
\includegraphics[width=0.46\textwidth, trim=0.2cm 0.4cm 0.cm 0.3cm,clip]{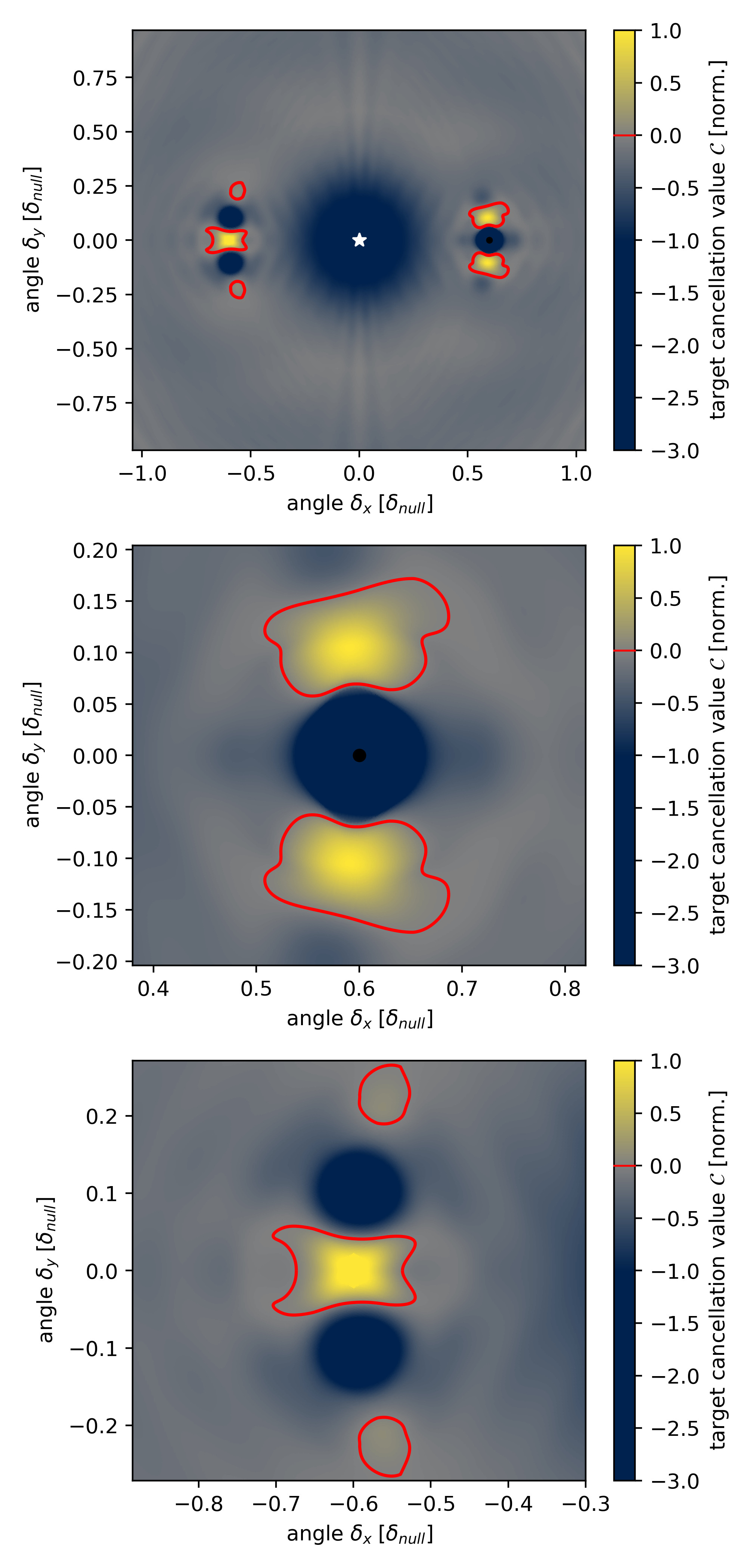}
  \caption{Top : Normalised target cancellation map $\mathcal{C}$ (Eq.\,\ref{TcancelM}) for $p_1 = ((0.6\cdot\delta_{\text{null}}, 0), \boldsymbol{F}_{p_1})$ (black dot) with homogeneous luminosity flux $\boldsymbol{F}_{p_1} = F_{p_1}\boldsymbol{1}$ across the full wavelength range, $\lambda$ from $4\,\mu$m to $18.5\,\mu$m and spectral resolution of 20. At each point ($\delta_x,\delta_y$) is computed the target cancellation map evaluated at $P = \{p_1,p_2\}$ where $p_2 = ( (\delta_x,\delta_y), \boldsymbol{F}_{p_2} = F_{p_1}\boldsymbol{1} )$ with constant noise independent from $P$, i.e.,$\sigma(\lambda)=\sigma$. When $p_2$ is outside the red region cancellation does not occur. The red region is independent from $F_{p_1}$ and $\sigma$. In this case $b_{\text{im}} = 6\,b_{\text{null}}$ and $\delta_0 \approx 0.13\,\delta_{\text{null}}$. (Middle) Zoom in the right region of Top plot. Bottom : Zoom in the left region of Top plot. }
  \vspace{-15pt}
     \label{photobombingmapcancel}
\end{figure}

We show (Fig.\,\ref{photobombingmapcancel}) the target cancellation map $\mathcal{C}$ (Eq.\,\ref{TCM}) for a configuration of two point sources with one point at a fixed position. We see that there are multiple regions of interest delimited by the 0 level line in red ($\mathcal{C}=0$). If $p_2$ is inside the region $\mathcal{C}>0$ then cancellation is possible. The resolution $\delta_0$ is defined using the area of this region. Notice that this is the worst case scenario where $p_1$ and $p_2$ have the same incoming luminosity flux. When the luminosity fluxes are different the cancellation disappears almost entirely. Indeed, if $F_1/F_2 = 0.5$ then for $p_1$ at $\delta_r = 0.6\,\delta_{\text{null}}$ there is no cancellation at all, i.e, $\mathcal{R} = 0$.
\subsection{Detection and contamination probabilities}
We want to calculate the probability of photobombing for a given planet knowing that it is detected. First, we define a detected planet to be a planet for which the integrated signal-to-noise ratio over all wavelengths for 100 hours integration time is above a threshold $\text{S}/\text{N}_{\text{target}} = \eta_{\text{S/N}}$. Thus, the detection indicator function is given by:
\begin{align}
    D_{p}(t) = H( \text{S/N}_{p(t)}(\Lambda) - \text{S}/\text{N}_{\text{target}} )
\end{align}
where $t$ represents the time evolution of the system, $p=p(t)$ describes the point source corresponding to the planet, $H$ is the Heaviside function, and $\text{S/N}_{p(t)}(\Lambda)$ is the signal-to-noise ratio over the wavelength range $\Lambda = [4\,\mu\text{m},18.5\,\mu\text{m}]$ after $100$ hours of integration time. This criterion pertains to the intensity of the signal rather than the final S/N. At each time $t$, we calculate the S/N of the planet as if we were able to integrate for 100 hours, with the planet remaining at a fixed position. According to \cite{Quanz_2022}, an integrated S/N of 7 is required for detection.\\ \\
We define the detection probability of a given planet/point source $p$ as the probability of detection. It is given by the following:
\begin{align}
    \mathbb{P}(D_{p} = 1) = P_D = \lim_{T\rightarrow \infty} \frac{1}{T} \int_{0}^{T} D_p(t) dt
\end{align}
for short observation times. \\ \\
We define the contamination indicator function of a point source/planet $p$ by checking if any other planet $p_1,p_2,\dots$ photobombs planet $p$. Its indicator function is given by:
\begin{equation}
    C_{p,P}(t) = H( \max_{i=1,2,\dots}{\mathcal{D}(\{p,p_i\},\Sigma)} )
\end{equation}
where $P = \{p_1,p_2,\dots\}$.\\ \\
From that, we define the contamination probability of a given planet/point source $p$ as the probability of contamination, knowing that it is detected. It is given by the following:
\begin{align}
    \mathbb{P}(C_{p,P}=1|D_p=1) = P_C = \lim_{T\rightarrow \infty}\frac{\int_{0}^{T} C_{p,P}(t)D_p(t) dt}{\int_{0}^{T} D_p(t) dt}
\end{align}
In practice, we approximate this quantity using Monte Carlo integration. Assuming circular orbits, we randomise the orbit angle $\theta$ of every planet uniformly over $[0,2\pi]$. For instance, a planet with a detection probability of 80\% means that it would be detectable approximately 80\% of the time, assuming the planet does not move along its orbit during an observation. Similarly, a contamination probability of 10\% means that 10\% of detections are contaminated,i.e., we detect only 1 planet instead of 2 planets. Note that the detection and contamination criteria do not take into account moving targets. 

The uncorrelation time is defined for a given planet and represents the time it takes for a first observation to become uncorrelated with a second observation, taking into account spectral contamination. We define the probability of having no contamination at time $t+T$ after contamination at time $t$ as $P_2(T)$. The uncorrelation time $T^\star$ is then the smallest time at which $P_2(T^\star) \approx P_C$. When the observation duration becomes comparable to the uncorrelation time, the assumption of static point sources is no longer valid. In such cases, considering the motion of planets during an observation becomes necessary to estimate whether contamination occurs. Unlike the contamination probability, the uncorrelation time captures information about the dynamics of the planets of the system we consider. An example illustrating this concept is presented in the Appendix D.

\section{Results and discussion} 
As mentioned earlier, there are several considerations one must bear in mind. Given two point sources/planets $p_1,p_2$ in the FOV with respective positions $\boldsymbol{\theta}_{p_1} = (\delta_{x,1}, \delta_{y,1}), \boldsymbol{\theta}_{p_2} = (\delta_{x,2}, \delta_{y,2})$ and incoming photon flux $\boldsymbol{F}_{p_1}, \boldsymbol{F}_{p_2}$ we can define a configuration $P = \{p_1, p_2\}$, 
\begin{itemize}
    \item The outputs of a given telescope are modelled as Gaussian multivariate distribution with mean related to the UIRF and covariance matrix related to the background noise.
    \item Photon noise is neglected, i.e., we assume the pixel size to be small compared to the variation of the UIRF.
    \item When $\mathcal{D}(P,\Sigma)>0 $ there is a possibility of contamination only. However, when $\mathcal{D}(P,\Sigma)<0$ there is no possible contamination. $\mathcal{D}$ is a proxy for a probability ratio, see (\ref{fraction_probabities}) for more details.
    \item $\delta_1$ works the same as the Rayleigh criterion $\delta_{\text{Rayleigh}}$. First, it overestimates spectral contamination, but $\delta_1$ depends on $\delta_x,\delta_y$ and the wavelength range $\Lambda$. Second, it generalises to more than two point sources, by checking contamination for each pair of point sources.
    \item During an observation, all planets are supposed to be at fixed positions. To better assess the potential of contamination in the case of moving targets, it is necessary to augment the space of images/time series generated, reconsider how to apply the parsimony principle, and find other numerical methods to solve the optimisation problem of finding $Q^\star$. 
    \item The contamination probability $P_C$ is defined by the probability that a given detected planet is contaminated, all while considering that the planet does not move during the observation. It is different from the probability of contamination during an observation $P_O(T)$ where time $T$ is the observation time. We always have $P_O > P_C$ and if $T <<$ uncorrelation time, we have $P_O \simeq P_C$. Both $P_C$ and $P_O$ are approximated using Monte Carlo integration.
\end{itemize}
\subsection{Solar system analogs}
In this section, we will approximate the detection and contamination probabilities using the following LIFE parameters (see Table \ref{tableLIFE}). They are the same as the ones used in \cite{Dannert_2022} for the Earth twin case. The nulling baseline will change according to the apparent HZ center angle and will be in the range of $[10\,\text{m}, 100\,\text{m}]$. Other parameters, such as stellar temperature, planet temperature, level of exozodi emission, etc., will be set by data from each individual system. 
\begin{table}[ht]
\caption{Fixed LIFE parameters for the singular stellar system studies. }              
\label{tableLIFE}      
\centering                                      
\begin{tabular}{c c c}          
\hline\hline                        
Parameter & Value & Description \\
\hline
$D$ &  2\,m & Aperture diameter \\
$R$ & 20 & Spectral resolution \\
$\eta_{\text{QE}}$ & 0.7 & Quantum efficiency \\
$\eta_t$ & 0.05 & Instrument throughput \\
$\text{S}/\text{N}_{\text{target}}$ & 7 & 100\,h S/N detection threshold \\
$\delta_H/\delta_{\text{null}}$ & 0.91 & apparent HZ center angle \\
$q$ & 6 & Scaling factor \\
\hline                                             
\end{tabular}
\end{table}
\subsubsection{Early type star system with long orbits}
\begin{figure}
    \centering
    \includegraphics[width=0.5\textwidth, trim=0.2cm 0cm 0.cm 0cm,clip]{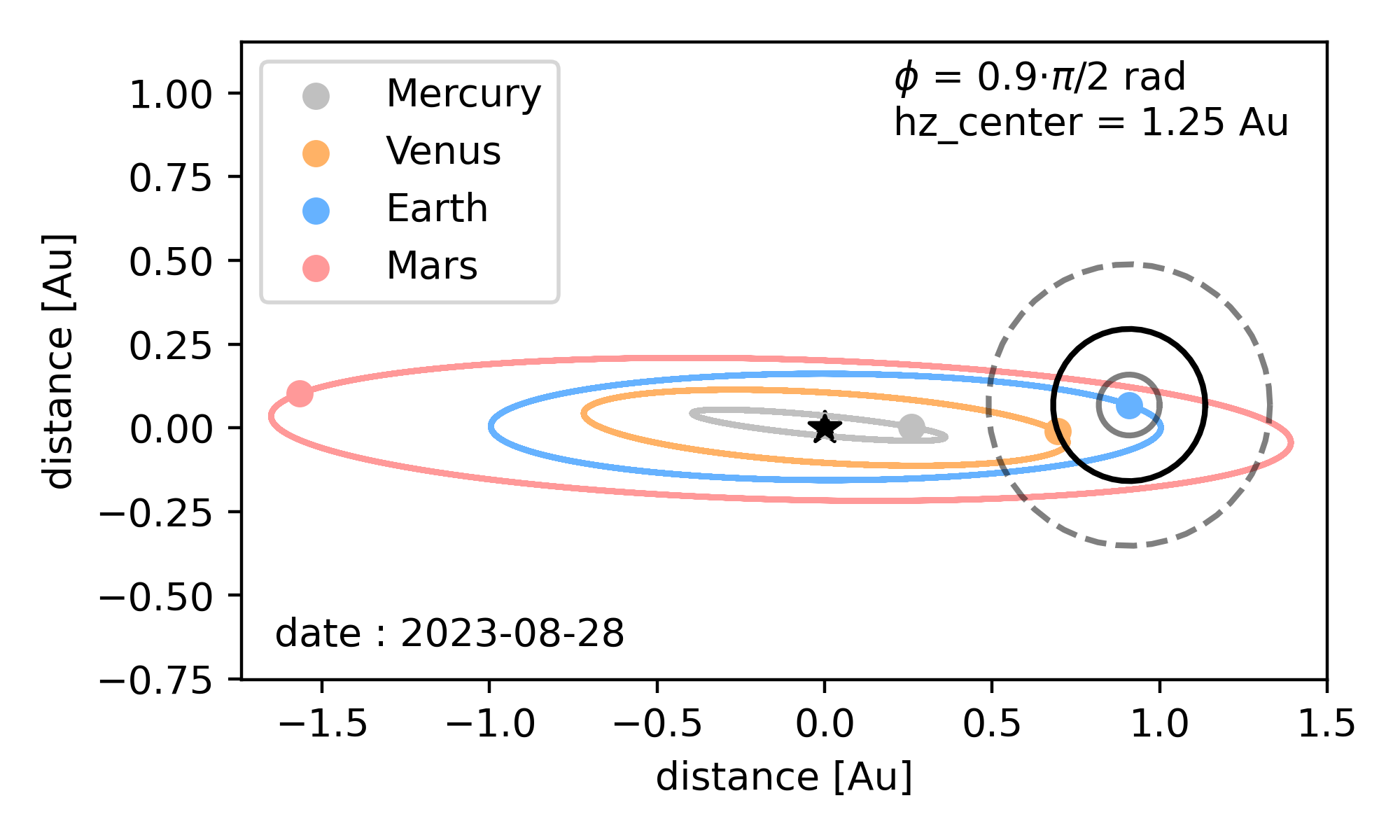}
\caption{Sun in the center, positions of Mercury, Venus, Earth and Mars at 28.08.2023 and trajectory of Mercury, Venus, Earth and Mars from 01.01.2020 to 01.01.2030 viewed from an angle of $\phi \approx 0.9\,\pi/2\,\text{rad}$ from the orbit plane of Earth. Circles represents the resolution of a traditional telescope with diameter $d = b_{\text{im}}$. The smaller (grey) for $\lambda = 4\,\mu\text{m}$, the medium (black) for $\lambda = 10\,\mu\text{m}$ and the large one (dashed grey) for $\lambda = 18.5\,\mu\text{m}$. The imaging baseline changes with the star system distance to have optimal sensitivity for planets in the HZ. This representation of the spatial resolution is valid for distances from 6.59 parsec ($60\,$m telescope Eq.) up to 65.9 parsec ($600\,$m telescope Eq.). Recall : this representation is not an accurate model to decide if photobombing occurs for LIFE. } \label{inner_solar_preview}
\end{figure}

\begin{figure}
    \centering
    \includegraphics[width=0.5\textwidth, trim=0.2cm 0cm 0.cm 0cm,clip]{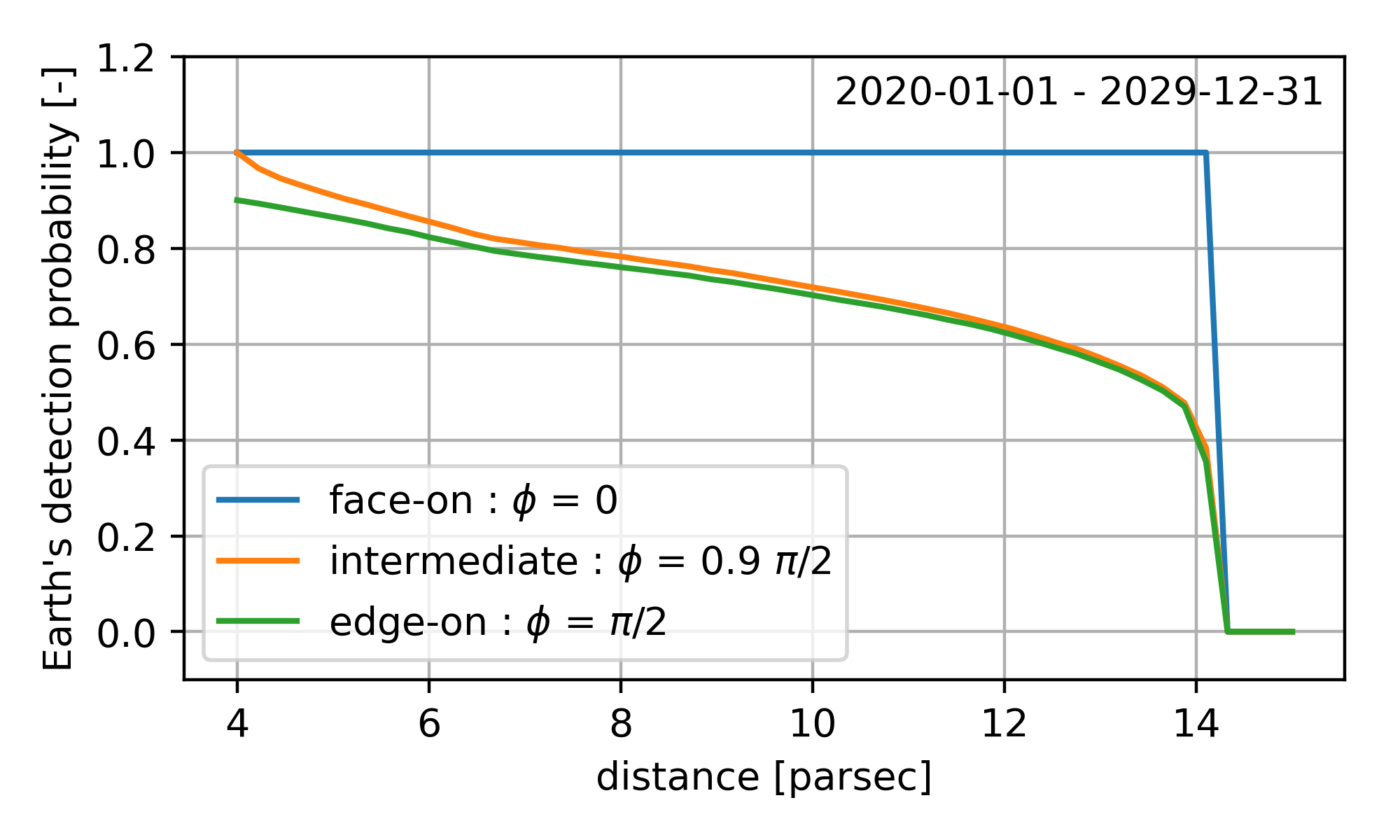}
\caption{The Earth's detection probability for multiple distances and three different inclination angles of the orbit plane for the period 2020-2030. The detection probability is the probability that the integrated S/N of Earth is above the threshold of 7 after an observation of 100 hours, where we assume that planets do not move. In the face-on case, the angular separation between Earth and the Sun is constant, thus Earth has the same S/N no matter its location in its orbit. When the system is at a far enough distance, the detection probability drops immediately from 1 to 0 because the S/N will not satisfy the detection criterion for the entire orbit at once.} \label{Earth_drate}
\end{figure}

Consider the solar system with Mercury, Venus, Earth, and Mars seen from a distance $d$ at an exozodi level of $3$. The position of each planet is calculated using Skyfield (\citeauthor{2019ascl.soft07024R} \citeyear{2019ascl.soft07024R}). We show (Fig.\,\ref{inner_solar_preview}) the apparent positions of orbits for a case between face-on and edge-on. In this case, the orbital plane of Earth is inclined by $\phi \approx 0.9\,\pi/2\,\text{rad}$.
The solar system can be seen with an optimal baseline, i.e., $\delta_{\text{HZ}} = \delta_H$, for distances $d \in$ [6.59 parsec, 65.9 parsec]. The Sun's spectrum is approximated by a black body spectrum of temperature $T=5778\,\text{K}$. The planet spectra are generated using the Planetary Spectrum Generator (PSG) developed by \cite{VILLANUEVA201886}. \\ \\
We show (Fig.\,\ref{Earth_drate}) the detection probability of Earth for multiple distances using an integrated S/N detection threshold of 7 for 100 hours of integration time. We note that for a distance above $\sim$14 parsecs, Earth would not be detectable under that criterion. Actually, since the detection criterion is relatively arbitrary, what matters more is how the probability of detection behaves. We understand that in a face-on scenario, the S/N will not depend on the position of Earth along its orbit, thus when the star distance increases, we get an abrupt decrease in the detection probability. Between 6.59 and 65.9 parsecs, the nulling baseline can be adjusted to ensure $\delta_H = \delta_{\text{HZ}}$, which maximises signal for a black body spectrum planet at $T=280\,\text{K}$ orbiting at the center of the HZ. In this particular case where the system is 10 parsecs away, we can set the nulling baseline appropriately such that we can ensure that $\delta_{0}$ matches $\delta_{\text{HZ}}$. The slow decrease of the edge-on case comes from the fact that parts of the orbit of Earth are very close to the center of the transmission map, yielding a lower probability of detection.

\begin{figure}
    \centering
    \includegraphics[width=0.5\textwidth, trim=0.2cm 0cm 0.cm 0cm,clip]{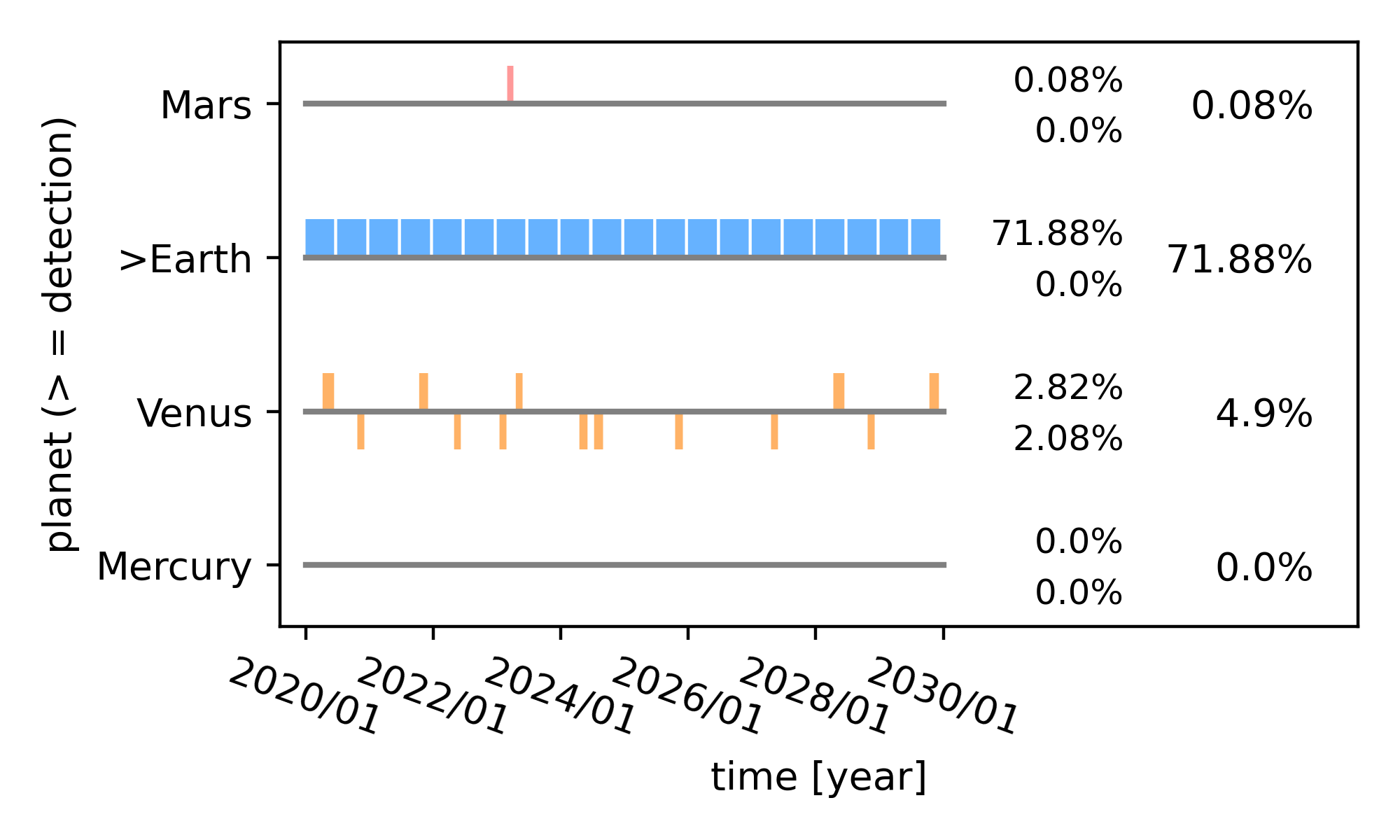}
\caption{ In blue over Earth's grey line, when Earth is detected in a case where $\phi\approx 0.9 \pi/2\,\text{rad} \approx 80\deg$ and star distance of $d = 10\,\text{parsec}$ over the period 2020-2030. Over the grey line (not Earth) when contamination is local and Earth detected. Under the grey line when contamination is non-local and Earth detected. Mercury does not contaminate Earth's spectrum at any time. Total contamination of 4.98\%. contamination probability of $0.0498/0.7188 \approx 7\%$. } \label{contamination_rate}
\end{figure}
\begin{figure}
    \centering
    \includegraphics[width=0.5\textwidth, trim=0.2cm 0.2cm 0.cm 0cm,clip]{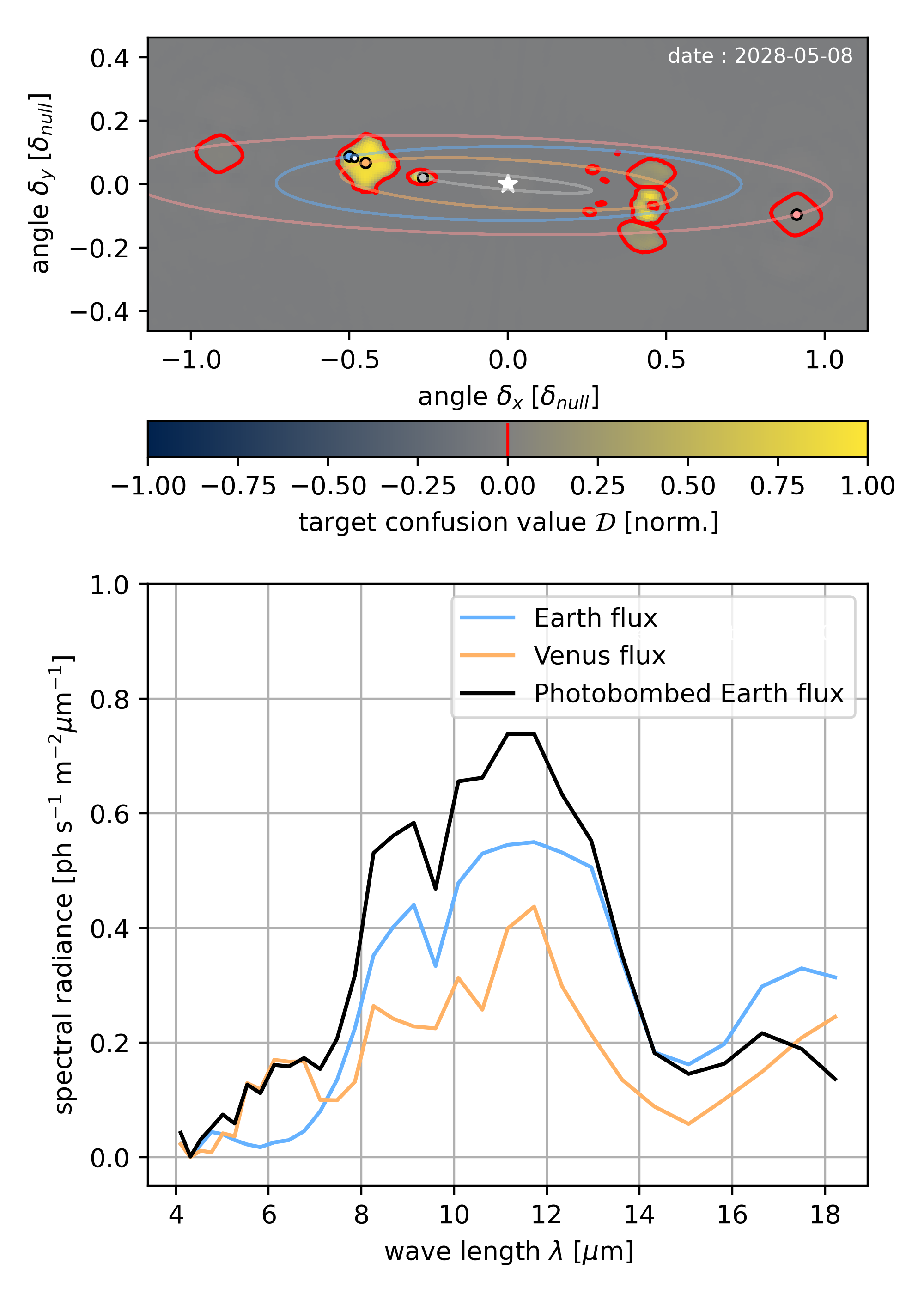}
\caption{  Top : Inner solar system seen from a near edge-on angle $\phi\approx0.9\pi/2\,\text{rad} = 81^\circ$. At each point $p = (\delta_x, \delta_y)$ is the maximum of the target confusion map between Earth at $p$ and another planet. Red level line is $\mathcal{D}=0$. Earth is inside a region where $\mathcal{D}>0$, i.e., Earth is photobombed. In this case Earth is photobombed by Venus. White dot represents the planet detected in that case. Bottom : Flux of Earth, Venus, and the Photobombed Earth. Photobombing depends on the flux of both Earth's and Venus's flux. The photobombed Earth's flux is the best point source that generates a signal similar to that of Earth and Venus together. } \label{EarthVenusPhotobomb}
\end{figure}
\begin{figure}
    \centering
    \includegraphics[width=0.5\textwidth, trim=0.2cm 0.2cm 0.cm 0cm,clip]{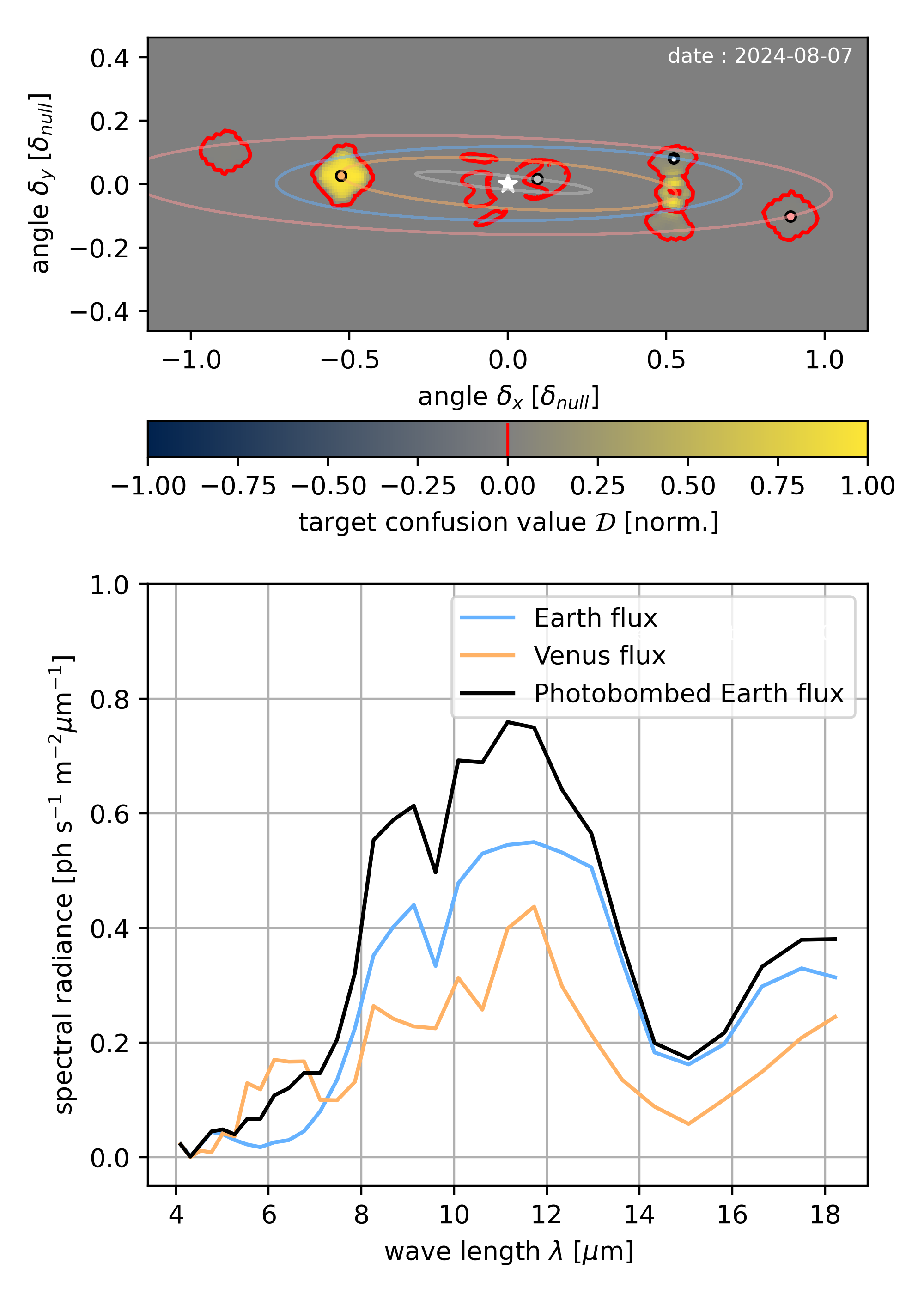}
\caption{  Top : Inner solar system seen from a near edge-on angle $\phi\approx0.9\pi/2\,\text{rad} = 81^\circ$. At each point $p = (\delta_x, \delta_y)$ is the maximum of the target confusion map between Earth at $p$ and another planet. Red level line is $\mathcal{D}=0$. Earth is inside a region where $\mathcal{D}>0$, i.e., Earth is photobombed. In this case Earth is photobombed by Venus non-locally. White dot represents the planet detected in that case. Bottom : Flux of Earth, Venus, and the Photobombed Earth. Photobombing depends on the flux of both Earth's and Venus's flux. The photobombed Earth's flux is the best point source that generates a signal similar to that of Earth and Venus together. } \label{EarthVenusPhotobombAnti}
\end{figure}

We show (Fig.\,\ref{contamination_rate}) the detection probability and contamination probability of Earth at 10 parsecs. The detection probability is approximately $72\%$. This means that $72\%$ of the time, Earth has a high enough signal intensity, defined by an S/N over 7, for a 100 hours observation, assuming Earth does not move. The contamination probability stands at approximately 7\%, indicating that in 7\% of detections, Earth has a contaminated spectrum by Venus or Mars. We notice that we recover the detection probability found in Fig.\,\ref{Earth_drate} at 10 parsecs with $\phi=0.9\,\pi/2\,\text{rad}$. Sometimes Earth gets contaminated by Venus locally (Fig.\,\ref{EarthVenusPhotobomb}) and other times it gets contaminated non-locally (Fig.\,\ref{EarthVenusPhotobombAnti}). Local contamination occurs when planets appear in close proximity, while non-local contamination arises when planets do not appear close to each other. In the case of LIFE, non-local contamination occurs when two planets are positioned on opposite sides. This is due to the symmetry of the transmission map. It is important to note that the photobombed Earth flux is not a simple sum or subtraction of the spectra of Earth and Venus. It is a more complex combination of the two. This is because the transmission map scales with the wavelength. At 10 parsecs, with $\phi \approx 0.9\pi/2\,\text{rad}$, the probability that Earth is contaminated during an observation of time $T_{\text{obs}}<100\,\text{hours}$ is close to the value of the contamination probability. This implies that the planets' movement is negligible for observations lasting less than 100 hours.
\subsubsection{Lower mass star system}
\begin{figure}
    \centering
    \includegraphics[width=0.5\textwidth, trim=0.4cm 0.2cm 0.2cm 0cm,clip]{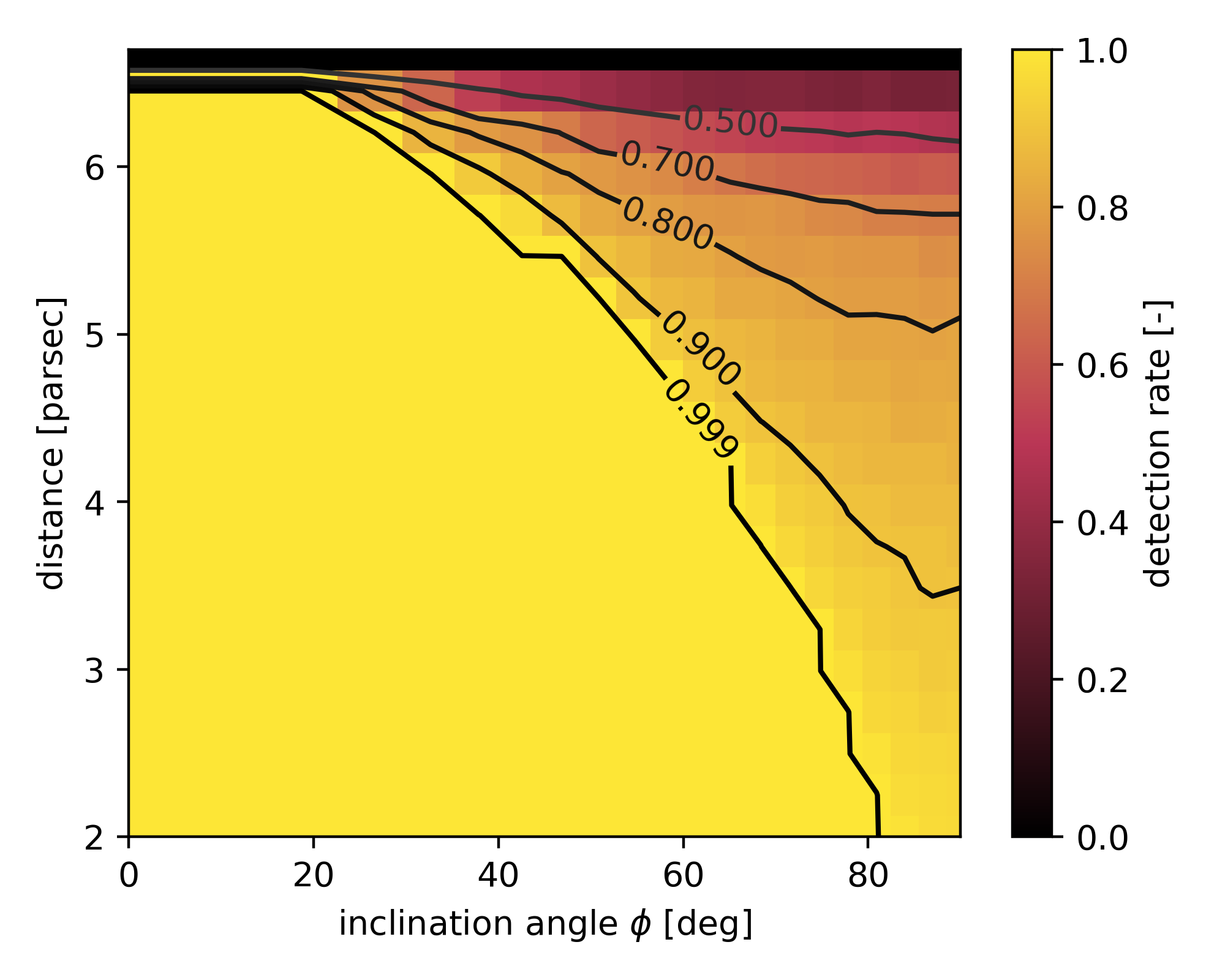}
\caption{  The detection probability of TRAPPIST-1e. The detection probability is the probability that the integrated S/N of TRAPPIST-1e is above threshold of 7 after an observation of 100 h. TRAPPIST-1e becomes undetectable for a distance above $\approx 6.5$ parsec under that criterion. } \label{TRAPPISTDrate}
\end{figure}

\begin{figure}
    \centering
    \includegraphics[width=0.5\textwidth, trim=0.4cm 0.2cm 0.2cm 0cm,clip]{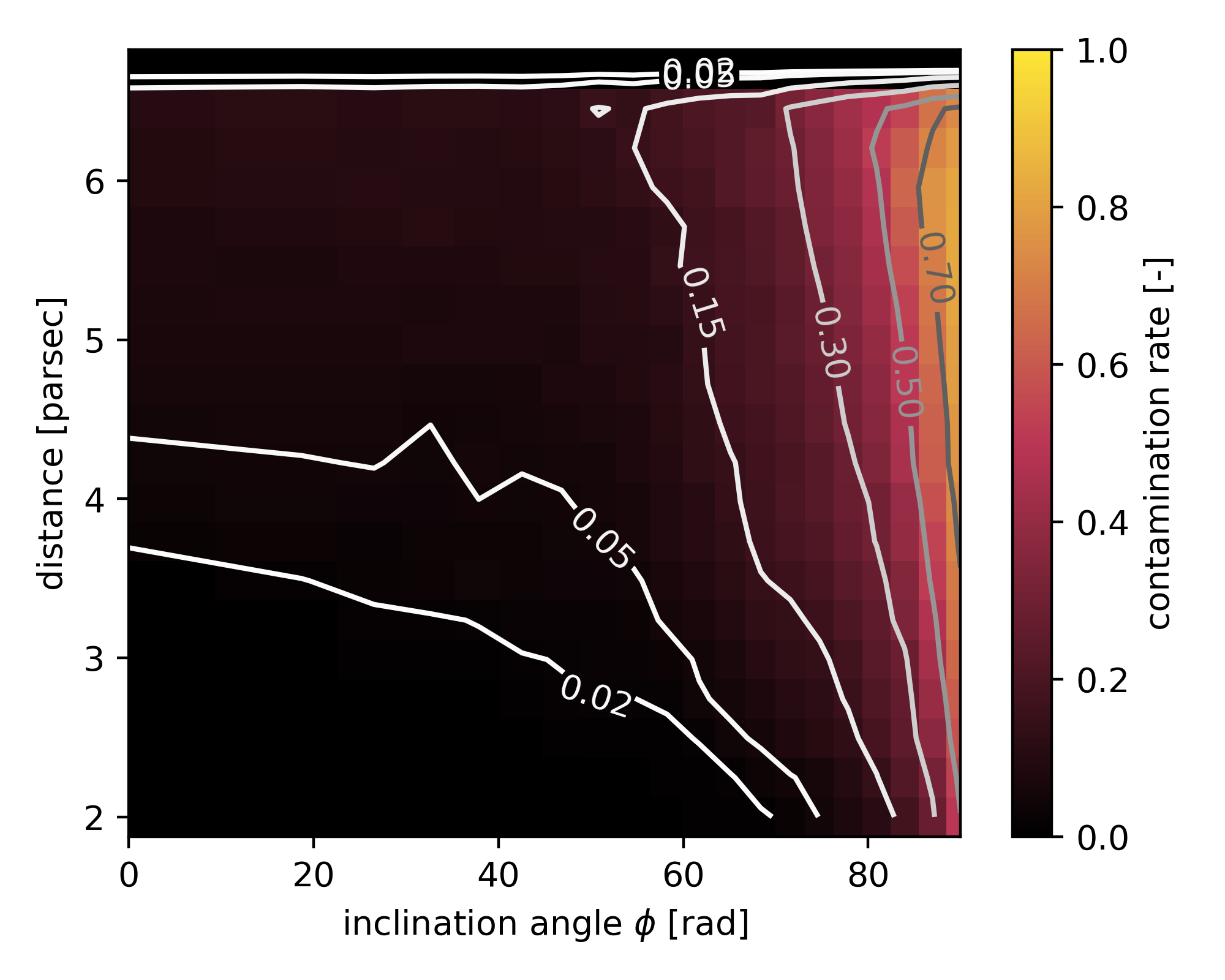}
\caption{  The contamination probability of TRAPPIST-1 e. The contamination probability is the probability that a given detected planet is contaminated. The contamination probability increase with distance and inclination angle. } \label{TRAPPISTCrate}
\end{figure}

In this section we study the TRAPPIST-1 system where we assume circular and coplanar orbits. We set the exozodi level to $3$ and show (Table\,\ref{tableTrapp}) relevant data used for the simulation. The planet's temperature allows us to generate a black body spectrum for each planet. Contrary to the previous case, we will calculate the detection/contamination probability using Monte-Carlo simulations. \\ \\
\begin{table}
 \caption{TRAPPIST-1 system data input. \cite{2017Natur.542..456G} contains all necessary features to generate simulations and estimate the contamination probability. More recent papers do not have all the necessary features. }
    \centering
    \begin{tabular}{c c c c}
    \hline
    \hline
& temp. [K] & radius [$\text{R}_{\odot}$] & HZ center [mAu] \\
\hline
\vspace{15 pt}
 a & 2559 & 0.12 & 35.68 \\
 \hline 
 \hline
& temp. [K] & radius [$\text{R}_{\oplus}$] &  semi-major [mAu] \\
\hline
b &   400.1 &     1.086 &      11.11 \\
c &   341.9 &     1.056 &      15.21 \\
d &   288.0 &     0.772 &      21.44 \\
e &   251.3 &     0.918 &      28.17 \\
f &   219.0 &     1.045 &      37.10 \\
g &   198.6 &     1.127 &      45.10 \\
h &   168.0 &     0.755 &      63.00 \\
\hline
\end{tabular}
 \label{tableTrapp}
\end{table}
Note that in this case the HZ center is relatively small and that we can maximise the expected signal for a black body planet of $\delta_H=\delta_{\text{HZ}}$ orbiting at the HZ center only if the distance is between $0.188\,\text{parsec}$ and $1.88\,\text{parsec}$. For larger distances the nulling baseline would be at maximum, i.e., 100\,m. We will focus our study on TRAPPIST-1e which has the most similar black body temperature to Earth. \\ \\
We show (Fig.\,\ref{TRAPPISTDrate}) the detection probability for multiple distances and inclination angles for planet TRAPPIST-1e. We notice the same behaviour as Earth for the face-on case ($\phi=0$), i.e., an abrupt fall of detection probability. Unlike the case of the solar system, at far distances the limiting factor is not the photon flux or surface area of detection but the nulling baseline, i.e., the size of the interferometer. The decrease in detection probability with the inclination angle is expected. This is because the planet has a higher likelihood of being situated closer to the center of the transmission map, thereby resulting in a weaker signal and in lower detection probability. \\ \\
We show (Fig.\,\ref{TRAPPISTCrate}) the contamination probability for multiple distances and inclination angles. Unlike the case of Earth, the contamination probability increases with distance. This is because we already are at the maximum nulling baseline. TRAPPIST-1e's apparent orbit approaches the center of the transmission map as the distance increases, whereas in the case of the solar system, the Earth remained at the same relative position to the transmission map. We observe that the contamination probability is strongly influenced by the inclination orbit angle, $\phi$. Even at identical distances, the contamination probability can vary significantly, ranging from less than 2\% to over 70\% for a distance of 3.5 parsec. Therefore, accurately determining the inclination angle using alternative methods is essential for estimating the extent of contamination. \\ \\
At 4 parsec, $\phi=0.9\pi/2\,\text{rad}$. The probability that TRAPPIST-1e is contaminated during an observation of time $T_{\text{obs}}>48\,\text{hours}$ is close to 1. This occurs because planets move relatively quickly in comparison to the observation time of LIFE. Similar to the previous section, we provide the uncorrelation time, which is approximately 15 hours. This duration is comparable to the typical observation time during the search phase, which lasts 10 hours up to a 100 hours. As mentioned previously, the $\delta_1$-resolution only considers fixed point sources. For these cases, we need to include moving point sources in the theoretical framework to determine whether two moving point sources are entangled or not.
\subsection{Population analysis}
The population analysis is done on simulated star systems. The data from \cite{Quanz_2022} contains 500 simulated universes with $\approx 1700$ star systems each, for which there is data about systems we can look at with LIFEsim. The distances range from 1 parsec to 20 parsec. LIFEsim allows us to calculate the background noise, signal intensity and planet noise for each planet in each system. We then use that to calculate the target confusion map and decide if contamination occurs. Note that in our study we have negligible planet noise compared to background noise.
\begin{figure}[ht]
\centering
\includegraphics[width=0.48\textwidth, trim=0.4cm 0.2cm 0.4cm 0cm,clip]{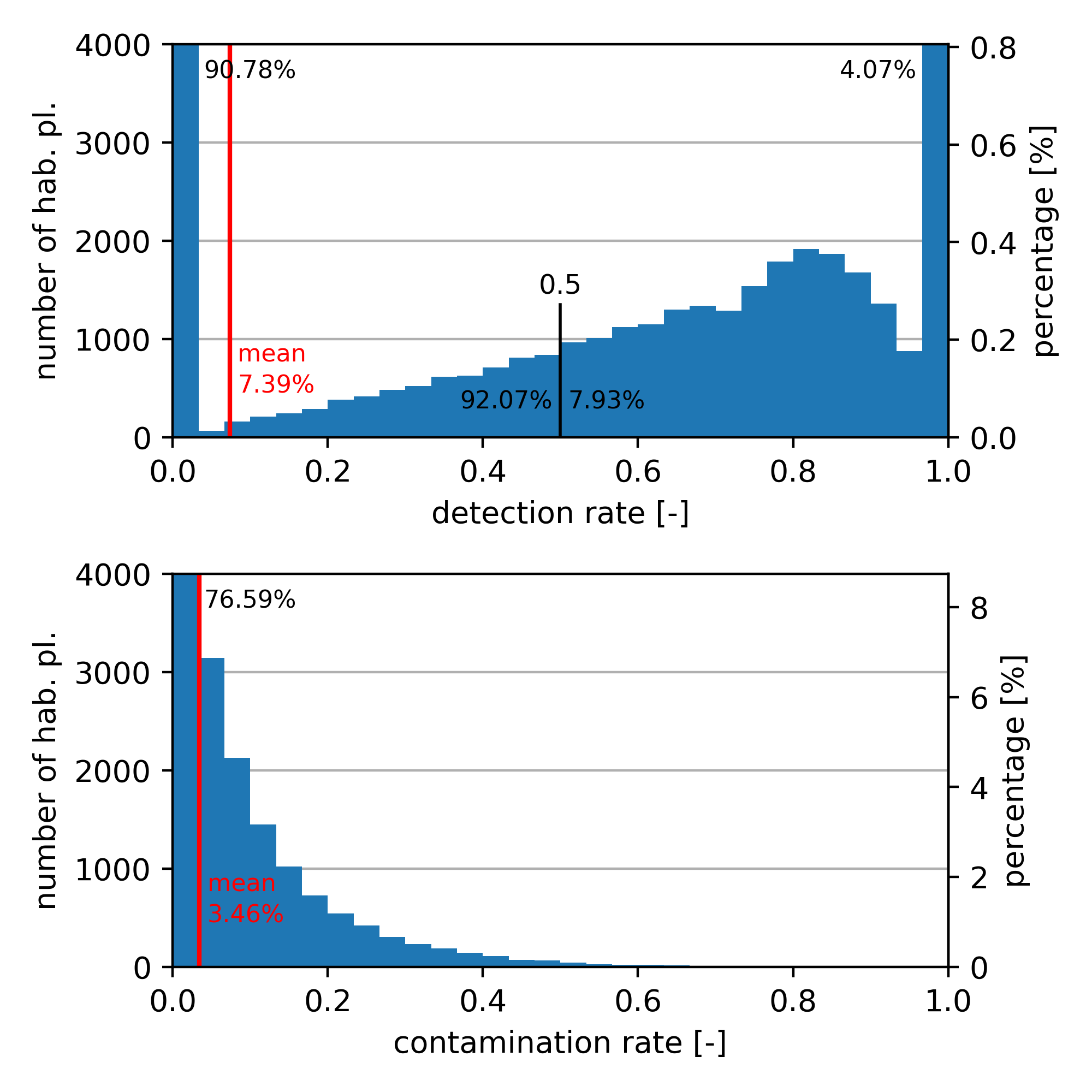}
\caption{Top : The detection probability for habitable planets for the entire LIFEsim survey (500 universes, $\approx$ 1700 systems each). The detection probability is the probability that a habitable planet has an S/N over 7 after 100 hours of observation, while assuming that the planet does not move during the observation. To better see the distribution, the left and right bars are cut off. They represent respectively 90.78\% and 4.07\% of planets. 7.93\% of the planets have a detection rate over 50\%. Bottom : The contamination probability for detectable habitable planets, i.e., detection probability $> 0\%$. The contamination probability is the probability of a detected planet to be contaminated by another planet in the same system. Similarly the first bar is cut off and represents 76.59\% of cases. } \label{cratedrateuniverses}
\end{figure}
\begin{table*}[ht]
\centering
\caption{The average number of habitable planet over universe samples (500 universes of $\approx$1700 stars each) and average number of habitable zone planets satisfying conditions on the detection probability and contamination probability. The detection criterion is to have, for a given habitable planet, an integrated S/N over wave length range $\Lambda=[4\,\mu\text{m},18.5\,\mu\text{m}]$ over $\text{S}/\text{N}_{\text{target}} =7$ for an integration time of 100 hours, while assuming that planets do not move during an observation. The detection probability is the probability of detection of a given habitable planet. The contamination probability is the probability that a given detected habitable planet is contaminated. The error is 1 standard deviation over universe samples. }
\begin{tabular}{lcccc}
\hline \hline 
  star type & M & K & G & F \\
\hline 
  \# = average number of habitable zone planets & $853 \pm 29$ & $115 \pm 11$ & $20.9 \pm 4.5$ & $3.8 \pm 1.8$ \\
  \# detection probability $>$ 0\%  & $78.1 \pm 8.3$ & $10.1 \pm 3.0$ & $2.5 \pm 1.7$ & $0.8 \pm 0.9$ \\
  \# detection probability $>$ 50\%  & $66.5 \pm 7.7$ & $9.2 \pm 2.9$ & $2.2 \pm 1.6$ & $0.8 \pm 0.8$ \\
  \# detection probability $>$ 50\% \& contamination probability $>$ 0\% & $24.3 \pm 5.0$ & $2.7 \pm 1.7$ & $0.7 \pm 0.9$ & $0.3 \pm 0.5$ \\
  \# detection probability $>$ 50\% \& contamination probability $>$ 5\% & $12.3 \pm 3.8$ & $1.2 \pm 1.1$ & $0.3 \pm 0.6$ & $0.1 \pm 0.3$ \\
  \hline
\end{tabular}
    \label{tabledetectableplanets}
\end{table*}
\\ \\
We show (Fig.\,\ref{cratedrateuniverses}) the detection and contamination probabilities of habitable zone planets for the entire database. About 8\% of habitable zone planets have a detection probability over $>50\%$. Moreover, we notice that $\approx 90\%$ are not detectable, meaning that at no point in time the planet will have an integrated S/N over 7 for a 100 hours observation. We show (Table\,\ref{tabledetectableplanets}) the average number of habitable zone planets with detection probability $>0\%$. These are refereed as detectable planets. We show the number of habitable zone planets with contamination probability $>0\%$. They represent $\approx 30\%$ of detectable planets. We show the number of habitable planets with contamination $>5\%$. They represent approximately 15\% of detectable planets. Lastly, considering that we observe each system for 100 hours, the expected number of detected habitable planets per universe is 73.4 (detected : $S/N_{\text{target}}>7$), and the expected number of detected and potentially contaminated habitable planets per universe is of 2.1. Thus, less than 2.8\% of habitable planets detections have contaminated spectrum. We remind the reader that the target confusion criterion overestimates contamination.
\section{Summary and conclusions}
This paper provides an approximate upper bound of spectral contamination occurrence of habitable zone planets by other planets in the same star system for LIFE. This is done by using the target confusion criterion described in Sect. \ref{specContam}. This notion is similar to how the Rayleigh criterion provides a higher estimate for contamination in traditional telescopes.  However, this concept can take into account the telescope's output as a whole, i.e., considering all observed wavelengths at once. For instance, LIFE offers the advantage of achieving a spatial $\delta_1$-resolution across its full wavelength range comparable to that of a very large traditional telescope of diameter $D = b_{\text{im}}= 6b_{\text{null}}$ observing at 4 $\mu$m, all while utilizing only four relatively small collector space craft. However, in the Double-Bracewell architecture, this benefit comes with trade-offs: first, we can observe only one system at a time; second, non-local contamination, meaning two planets may appear as one even if they are not in close proximity; and third, cancellation of point sources can occur, meaning that two point sources can cancel each other's signal while being detectable independently. Non-local contamination arises from the antisymmetric nature of LIFE's transmission map. Cancellation occurs due to the destructive interference of signals in an interferometer.

In the detection phase, LIFE will perform a survey of nearby stars in search for habitable and potentially inhabited planets. These observations need typical timescales of tens of hours. For the actual characterisation phase of LIFE, the timescales of observations are even longer. From \cite{Konrad_2022} and \cite{2022A&A...665A.106A}, we learned that the characterisation of Earth twins at 10 parsecs will take on the order of 50-100 days, while, for example, \cite{2023AsBio..23..183A} showed that the detection of potentially biogenic gases in the atmosphere of later-type systems will take on the order of tens of days for targets at typical distances of around 5 parsecs. The target confusion criterion used in this paper assumes that planets do not move during an observation, again similar to how the Rayleigh criterion works. Incorporating moving point sources would further complicate the theoretical framework by expanding the space of possible signals that moving planets can produce. Thus, this criterion to decide if contamination occurs is accurate when the uncorrelation time is much greater than the observation time. This is the case only for early-type stars and observations that are shorter than 100 hours, which is typical during the search phase.

To summarise, lower mass stars like TRAPPIST-1 illustrate a resolution limitation in LIFE, leading a priori to more frequent contamination. This limitation arises because the nulling baseline cannot satisfy optimal signal intensity conditions, i.e., $\delta_H = \delta_{\text{HZ}}$. For star distances greater than 2 parsecs, the nulling baseline would be fixed at its maximum value of 100 meters, bringing the apparent HZ closer to the center of the transmission map. Conversely, early-type stars like the Sun highlight a sensitivity limitation. While the nulling baseline may be optimal for larger distances, there comes a point where, for instance, Earth would no longer be detectable. In this case, the contamination probability does not vary significantly with star distance because the HZ would scale with the transmission map. In both extreme scenarios, the contamination probability is significantly influenced by the orbital inclination angle and the number of planets. For example, for TRAPPIST-1 at $\sim$ 3.5 parsecs, the contamination probability can range from under 2\% in the best case (face-on) to more than 70\% in the worst case (edge-on). For Earth, almost all contamination comes from Venus. In the simulation, Venus is responsible of approximately 98\% of the contamination of Earth. We saw in the inner Solar system example that LIFE can be prone to non-local contamination, i.e., where planets with diametrically opposite positions in the field of view can appear as one planet. Moreover, we saw in that example that this "one planet" has a spectrum that is not a simple linear combination of the spectra of Earth and Venus, but a more complex combination of the two. The population analysis based on the search campaign simulations carried out by \cite{Quanz_2022} showed that only 71.3 out of 73.4 detected habitable zone planets are not contaminated on average. Hence, this study shows that spectral contamination is not a severe problem for LIFE. The simulations showed that, after all, the increase in spectral contamination for M-type star systems is not significant compared to G-type star systems. We note that this study does not account for other non-planet sources, such as zodiacal clouds, which can resemble planets and contaminate the signal, occurring a priori more frequently in edge-on cases. One approach to addressing this could be incorporating them into higher-fidelity simulations. This issue must be considered to better quantify the frequency and extent of spectral contamination.

A potential solution to mitigate contamination when the observation time is short compared to the uncorrelation time may be to observe multiple times and wait between observations for the system to uncorrelate, i.e., to not look the same. This will increase the chance of having at least one non-contaminated sample. However, when the observation time is comparable or longer than the uncorrelation time, this technique may not be effective anymore because the signal would not necessarily be contaminated or non-contaminated during the entire observation window. In that case, we can only integrate the incoming signal until we are confident enough about the configuration of planets in the system. This amounts to relying on the second condition (Eq. \ref{condition_probable}), i.e., having a sufficiently high S/N, to reduce the probability that one virtual planet can produce the images/timeseries produced by two planets. To decide if contamination occurs for longer observations, it is necessary to extend the theoretical framework to accommodate moving targets by expanding the space of images/timeseries to include those produced by two moving point sources and reconsider the application of the parsimony principle. Then, finding the best configuration of moving point sources in the much larger space of images/timeseries. This optimisation problem may require different techniques than those used in this paper. Assuming Keplerian dynamics, as done in \cite{refId0}, or having additional information about the system from other observation sites can help solve the optimization problem more efficiently. Another potential way to mitigate spectral contamination is to set up different combination schemes to reduce or eliminate degeneracies, such as non-local contamination.

Lastly, the new criteria for contamination and cancellation may also be used in studies investigating different beam combination schemes and mission architectures (\citeauthor{Hansen_2022}). Indeed, one could envision finding the best transmission map in a given transmission map space to minimise the resolution $\delta_1$ or the average number of contaminated planets for a given star population. That is not the sole measure, but it can help guide us toward better nulling interferometer architectures.

\begin{acknowledgements}
    The material is based upon work supported by NASA under award number 80GSFC24M0006. The authors also acknowledge support from the Goddard Space Flight Center (GSFC) Sellers Exoplanet Environments Collaboration (SEEC), which is supported by the NASA Planetary Science Division’s Research Program. D.A.'s work has been carried out within the framework of the National Centre of Competence in Research PlanetS supported by the Swiss National Science Foundation under grants 51NF40\_182901 and 51NF40\_205606. This work has received funding from the Research Foundation -  Flanders (FWO) under the grant number 1234224N. 
\end{acknowledgements}

%


\software{LIFEsim \citep{Dannert_2022}, PSG \citep{VILLANUEVA201886} ,Sympy \citep{10.7717/peerj-cs.103}, Skyfield \citep{2019ascl.soft07024R}}


\newpage
\appendix
\section{Numerical method}
In the generalisation of spatial resolution, we must find a configuration $Q$ containing a single point source that exhibits a photon density distribution most similar to that of configuration $P$ containing two point sources. This result will allow to compute the target confusion $\mathcal{D}(P,\Sigma)$. The method have to find $\boldsymbol{\theta}$ the angular position and $\boldsymbol{F}$ the flux. We update these parameters via the Newton Raphson method. The gradient and the hessian are given by the following, 
\begin{align*}
     \nabla_{\boldsymbol{\theta},i} &:= 2 (\boldsymbol{\mu}-\boldsymbol{d})^t \Sigma^{-1}\partial_i\boldsymbol{\mu} \\ 
     &=2\text{sum}\left((\Sigma_F^{-t} M_{\boldsymbol{\mu}-\boldsymbol{d}}^t \Sigma_G^{-1})^t \circ M_{\partial_i\boldsymbol{\mu}}\right) \\
     H_{\boldsymbol{\theta},ij} &:= 2 \left( \partial_i\boldsymbol{\mu}^t\Sigma^{-1}\partial_j\boldsymbol{\mu} + (\boldsymbol{\mu}-\boldsymbol{d})^t \Sigma^{-1} \partial_i\partial_j\boldsymbol{\mu}\right) \\ 
    &=2\text{sum}\left(M_{\partial_i\boldsymbol{\mu}^t} \circ (\Sigma_G^{-1} M_{\partial_j \boldsymbol{\mu}} \Sigma_F^{-t}) + (\Sigma_F^{-t} M_{\boldsymbol{\mu}-\boldsymbol{d}}^t \Sigma_G^{-1})^t \circ M_{\partial_i\partial_j\boldsymbol{\mu}} \right)  \\
    &\theta \leftarrow \theta - H_{\boldsymbol{\theta}}^{-1}\nabla_{\boldsymbol{\theta}}
\end{align*}
where $\boldsymbol{\mu} = \text{vec}(M_{\boldsymbol{\mu}})$, $\circ$ is the Hadamard product and $\text{sum}$ is the element wise sum. The fluxes $F$ are updated via linear regression. It is an approximation from using Newton Raphson on all the parameters, i.e., $\boldsymbol{\theta}$ and $\boldsymbol{F}$.

Remark: With this method, the global minimum $Q^{\star}$ is not guaranteed; however, we can assume that when $p_1$ and $p_2$ are about to contaminate each other, the solution $Q^{\star}$ will be in the neighbourhood of one of them. This comes from the smoothness of the cost function with respect to the position of each $p_1$  and $p_2$ and the fact that we are interested when the total signal looks like the signal of a single point source.
\section{Varying $\alpha$ for a traditional telescope}
\begin{figure}[ht]
    \centering
    \includegraphics[width=0.5\textwidth, trim=0.4cm 0.2cm 0.2cm 0cm,clip]{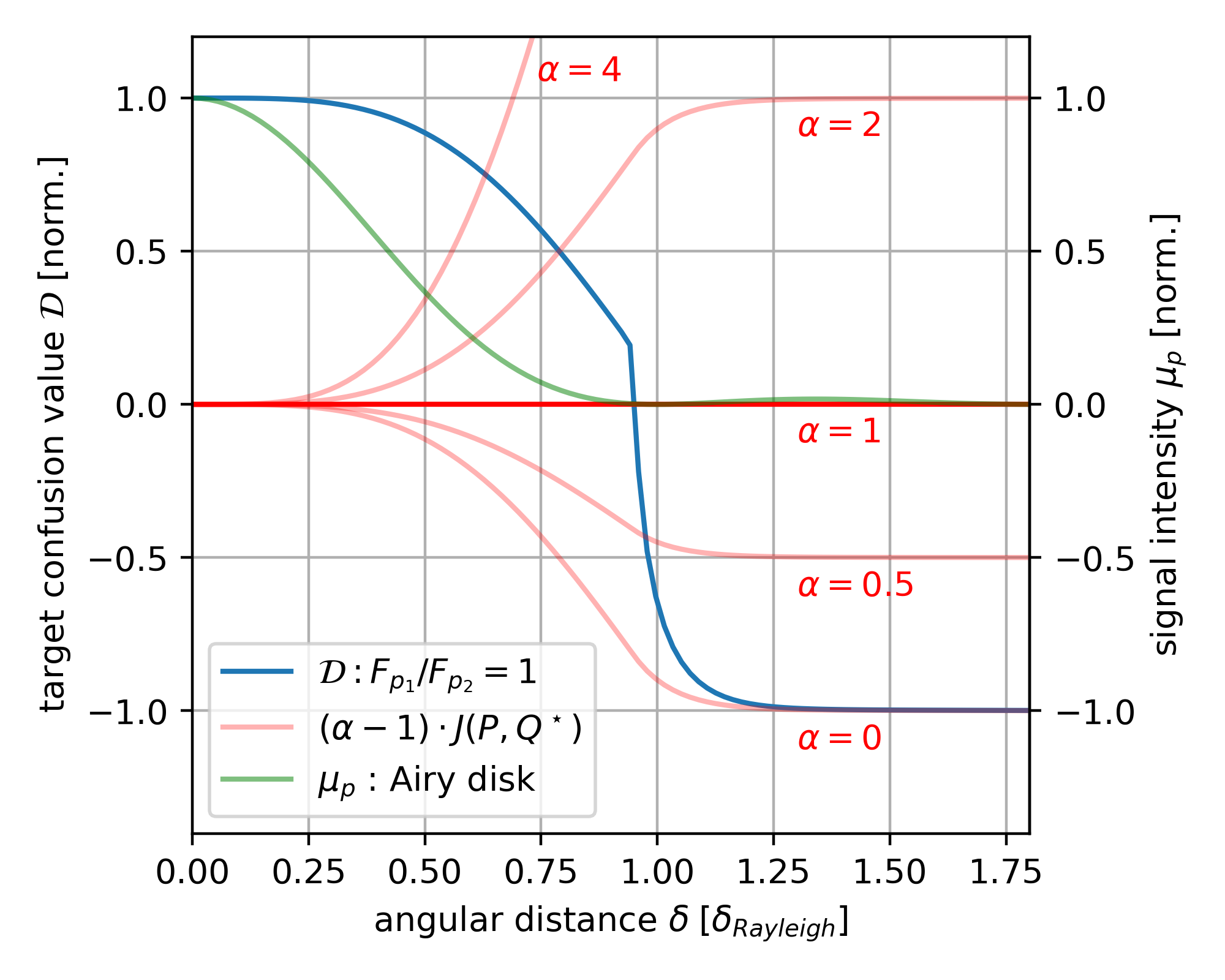}
\caption{ Target confusion value $\mathcal{D}$ (blue) compared to the RHS term in (Eq.\,\ref{condition_different}) (red) for different values of $\alpha$. If $\mathcal{D}(P,\Sigma) < (\alpha-1)J(P,Q^\star,\Sigma)$ then by definition of $\alpha$ there is no contamination.} \label{Airyapproach3}
\end{figure}

We show (Fig.\,\ref{Airyapproach3}) how the condition (\ref{condition_different}) is satisfied for different values of $\alpha$. We see that the greater $\alpha$ is, the weaker the contamination criterion becomes. In the paper we choose $\alpha = 1$.
\section{Convergence rate of $\delta_1$-resolution study}
\begin{figure}[ht]
    \centering
    \includegraphics[width=0.48\textwidth, trim=0.4cm 0.2cm 0.cm 0cm,clip]{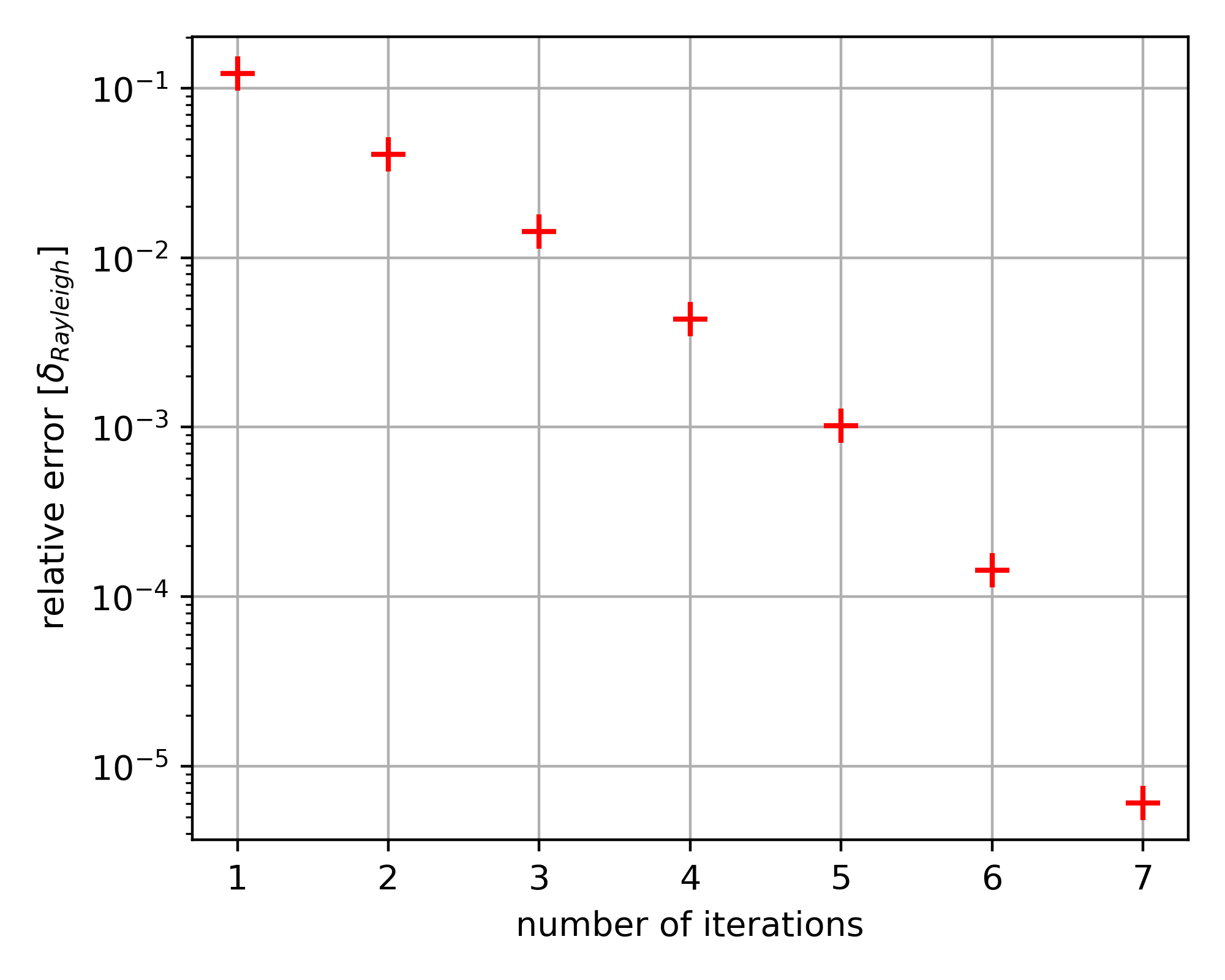}
\caption{   Error $\delta_1(\{\lambda\})- \delta_{\text{C},11}$ in the number of iterations. where $\delta_{\text{C},11}$ is $\delta_1(\{\lambda\})$ computed at 11 iterations. 1 iteration correspond to step 2 and 3 in the numerical method described in Sect. 2.6.} \label{convergence_Airy}
\end{figure}

We show (Fig.\,\ref{convergence_Airy}) the convergence rate of the $\delta_1$-resolution of traditional telescopes using the numerical method described in Sect. 2.6. Since there is no known analytical solutions we compared $\delta_1$ to the solution found for a higher number of iterations, here 11. We see that the convergence rate is exponential at least.

\section{Uncorrelation time example}
\begin{figure}[ht]
    \centering
    \includegraphics[width=0.5\textwidth, trim=0.1cm 0.1cm 0.1cm 0cm,clip]{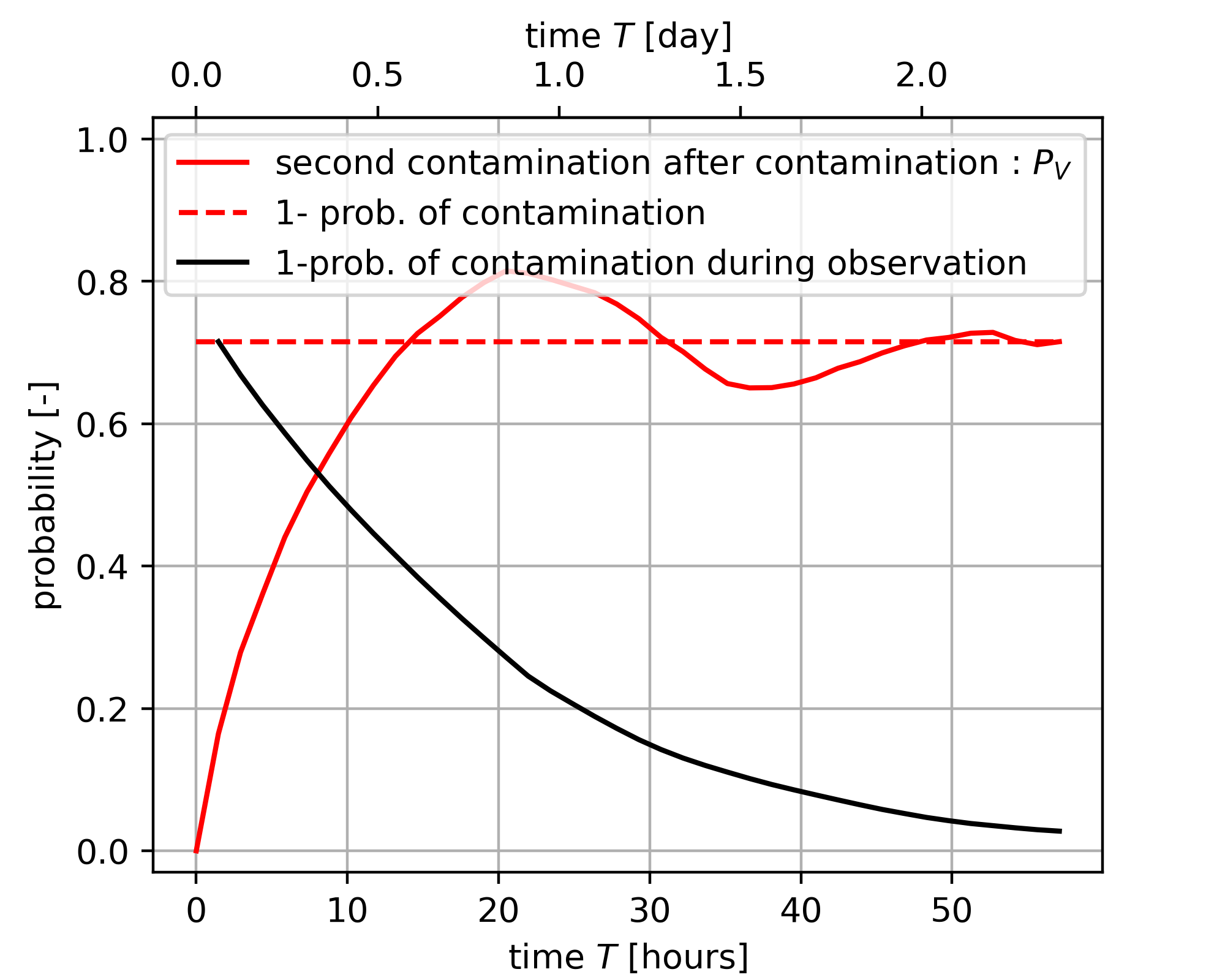}
\caption{  In red is the probability $1-P_2(T)$ to have contamination at time $t+T$ after having contamination on the first visit at time $t$ for TRAPPIST-1e. Dashed red is $1-P_C$ where $P_C$ is the contamination probability of TRAPPIST-1e. In black is the probability $1-P_O(T)$ of no contamination during an observation of time $T$. We read that the uncorrelation time is of 15 hours. We notice that $1-P_O(T>2 \mbox{days}) \approx0$. Hence, in this case considering that the system does not change for observations > 15 hours is a bad approximation.} \label{secondvisitTRAPPISTSystem}
\end{figure}

In this section we show how the uncorrelation time is deduced. Given a timeseries of the contamination indicator function of a planet within a configuration of planets we can calculate the probability $P_2(T)$ of having no contamination at time $t+T$ after contamination at time $t$, using Monte Carlo integration techniques. We show (Fig.\,\ref{secondvisitTRAPPISTSystem}) that probability for the case of TRAPPIST-1e in the TRAPPIST-1 system. 
\section{Variables Index}
\begin{table}[ht]
    \centering
    \caption{Index of symbols used in this paper.}
    \begin{tabular}{lp{13.8cm}}
    \hline
    \hline
     symbol &  description \\
    \hline
     $\delta $ &  apparent angle \\
     $S_{\text{dist}}$ & star distance \\
     $r$ &  apparent distance, $r= \delta \cdot S_{\text{dist}}$. \\
     $\lambda$ & wavelength \\
     $\Lambda$ & wavelength range, $\Lambda = [4\,\mu\text{m},18.5\,\mu\text{m}]$ \\
     $\phi$ &  orbit plane inclination (face-on $\phi=0\,\text{rad}$, edge-on $\phi=\pi/2\,\text{rad}$)  \\
     $\theta$ & orbit angle (right side $\theta=0$), used when orbit assumed to be circular  \\
     $b_{\text{null}}$ & null baseline, ranges from 10 m to 100 m for LIFE \\
     $q$ & scaling factor, $q=6$ for LIFE \\
     $T$ & transmission map   \\
     $S$ & sensitivity map  \\
     $p=(\delta_x, \delta_y, l)$ & point source at ($\delta_x,\delta_y$) with incoming luminosity flux $l$  \\
     $U_p$ & Unit instrument response function, function that describes the noiseless output signal of a given point source $p$ with unit flux luminosity in the FOV, same as PSF idea. In this paper it can be an image or a timeseries. \\ 
     $\mu_p$ & function that describes the noiseless output signal of a given point source $p$ in the FOV. In this paper it can be an image or a timeseries. \\
     $\sigma$ & total noise term (photon noise = Poisson noise $\approx$ gaussian noise) coming from background noise $\sigma_b$ (stellar leakage, local zodiacal dust, exozodiacal dust) and planet noise $\sigma$. \\
     $P=\{p_1,\dots,p_N\}$ & set or configuration of point sources \\
     $J(P,Q,\Sigma)$ & cost function between timeseries of two configuration of point sources $P, Q$ with background noise $\Sigma$. \\
    $\mathcal{D}(P, \Sigma)$ & target confusion map. If $\mathcal{D}<0$ then there is no possible contamination. \\
    $\mathcal{C}(P,\Sigma)$ & target cancellation map. If $\mathcal{C}<0$ then there is no possible cancellation. \\
    $\mathcal{R}(p_1,\boldsymbol{F}_{p_2},\Sigma)$ & resolution map, fix 1 point source and move around the other. Is the area of the positions of $p_2$ where a given criterion is satisfied. $\mathcal{D}(\{p_1,p_2\},\Sigma)>0$ for contamination and $\mathcal{C}(\{p_1,p_2\},\Sigma)>0$ for cancellation. \\
    $\delta_{0},\delta_1, \delta_{\text{C}}$ & generalised spatial resolution for the criteria $\mathcal{D}>0$,$\mathcal{C}>0$ and $\mathcal{D}>0\cup \mathcal{C}>0$ respectively. Equals $\mathcal{R}$ for the case where $\boldsymbol{F}_{p_1} = \boldsymbol{F}_{p_2} = F\boldsymbol{1}$ for $F>0$ and $\Sigma = \sigma \boldsymbol{1}$ for $\sigma>0$ \\ 
    $D_{p}(t)$ & detection indicator function. is 1 if the S/N of 7 is reached in less than 100 h, assumed planet stay at a fixed position.\\
    $C_{p,P}(t)$ & contamination indicator function, detection and contamination criterion $\mathcal{D}(P,\Sigma)>0$ satisfied.\\
    $P_D$ & detection probability, probability that a given planet is detected. \\
    $P_C$ & contamination probability, probability that a given detected planet is contaminated. \\
    $T^\star$ & uncorrelation time, smallest time between two observations to have uncorrelated contamination occurrence.\\
    \hline
    \end{tabular}
    \label{variables_index}
\end{table}
\newpage
\bibliography{sample631}{}
\bibliographystyle{aasjournal}



\end{document}